\documentclass[11pt]{article}
\pdfoutput=1
\usepackage{CLICdp}
\usepackage{epstopdf}
\usepackage{graphicx}

\def\pt{\ensuremath{p_{\mathrm{T}}}} 

\def\TeV{\ifmmode {\mathrm{\ Te\kern -0.1em V}}\else
                   \textrm{Te\kern -0.1em V}\fi}%
\def\GeV{\ifmmode {\mathrm{\ Ge\kern -0.1em V}}\else
                   \textrm{Ge\kern -0.1em V}\fi}%
\def\MeV{\ifmmode {\mathrm{\ Me\kern -0.1em V}}\else
                   \textrm{Me\kern -0.1em V}\fi}%
\def\keV{\ifmmode {\mathrm{\ ke\kern -0.1em V}}\else
                   \textrm{ke\kern -0.1em V}\fi}%
\def\eV{\ifmmode  {\mathrm{\ e\kern -0.1em V}}\else
                   \textrm{e\kern -0.1em V}\fi}%
\let\tev=\TeV
\let\gev=\GeV

\title{Jet reconstruction at high-energy electron-positron colliders}

\clicdppub{2017}{002}
\date{\today}

\addauthor{M.~Boronat}{\institute{1}}
\addauthor{J.~Fuster}{\institute{1}}
\addauthor{I.~Garcia}{\institute{1}\thanks{Corresponding author: ignacio.garcia@ific.uv.es}}
\addauthor{Ph.~Roloff}{\institute{2}}
\addauthor{R.~Simoniello}{\institute{2}}
\addauthor{M.~Vos}{\institute{1}}
\addinstitute{1}{IFIC (CSIC/UVEG), Valencia, Spain}
\addinstitute{2}{CERN, Geneva, Switzerland}

\abstract{
In this paper we study the performance in $e^+e^-$ collisions of classical 
$e^+e^-$ jet reconstruction algorithms, longitudinally invariant algorithms 
and the recently-proposed Valencia algorithm. 
The study includes a comparison of 
perturbative and non-perturbative jet energy corrections and the response
under realistic background conditions. Several algorithms are benchmarked
with a detailed detector simulation at $\sqrt{s}= 3$~\tev{}.
We find that the classical $e^+e^-$ algorithms, with or without beam jets, 
have the best response, but are inadequate in environments with non-negligible
background. The Valencia algorithm and longitudinally 
invariant $k_t$ algorithms have a much more robust performance, with a slight 
advantage for the former.
}
\titlecomment{This work was carried out in the framework of the CLICdp collaboration}
\begin{document} 
\titlepage


\section{Introduction}
\label{sec:intro}


The next large collider facility could be a high-energy electron-positron 
collider. Linear $e^+e^-$ colliders are the best tool to 
explore the energy range from several 100~\gev{} to a few \tev.
The Technical Design Report of the International Linear 
Collider (ILC~\cite{Fujii:2015jha,Baer:2013cma}) project envisages a 
programme of precision Higgs and top physics
at centre-of-mass energies $\sqrt{s}=$ 250~\gev, 500~\gev{} and, 
after an energy upgrade, 1~\tev. 
The Compact Linear Collider (CLIC~\cite{Linssen:2012hp})
scheme has been shown to reach accelerating gradients that
extend the $e^+e^-$ programme into the multi-\tev{} regime. CLIC 
envisages a first phase at $\sqrt{s}=$   
380~\gev{}, followed by its high-energy programme of multi-\tev{} operation,
with stages at 1.5~\tev{} and 3~\tev{}~\cite{CLIC:2016zwp}. 
A large circular machine, as envisaged by the 
FCCee~\cite{Gomez-Ceballos:2013zzn} and 
CEPC~\cite{CEPC-SPPCStudyGroup:2015csa} projects, could provide
high luminosity at 250~\gev{}. FCCee may reach the top quark pair
production threshold~\cite{Gomez-Ceballos:2013zzn}.
On a longer time scale a muon collider~\cite{Alexahin:2013ojp} 
also has the potential to reach the multi-\tev{} regime.

Measurements of hadronic final states are a key ingredient of the 
programme of any next-generation lepton
collider. Excellent jet reconstruction is essential
to characterize the couplings of the Higgs boson and top quark at
the sub-percent level. To distinguish hadronic $W$ and $Z$ boson decays
a jet energy resolution of approximately 3\% is required. 
The linear collider detector 
concepts~\cite{Behnke:2013lya,AlipourTehrani:2254048} achieve
this performance with highly granular 
calorimeters~\cite{collaboration:2010rq,Adloff:2012gv} 
and particle-flow algorithms~\cite{Marshall:2015rfa}.
Excellent jet clustering is needed to benefit fully from their potential.


The increase in energy comes with a number of challenges for jet
reconstruction. Compared to LEP and SLC, high-energy machines 
produce an abundance of 
multi-jet final states, final states with multiple energy scales (in associated 
production), forward-peaked processes and highly boosted objects. 
Backgrounds such as 
$\gamma \gamma \rightarrow $ {\em hadrons} production
are increasingly important at high 
energy~\cite{Chen:1993dba}.
The classical $e^+e^-$ algorithms cannot cope with this environment,
 in particular with the beam-induced  
background~\cite{Linssen:2012hp,Marshall:2012ry,Simon:2015pza}.
A critical evaluation of jet reconstruction at lepton colliders is
therefore mandatory.

In this paper we study the performance of jet reconstruction in multi-\tev{}
$e^+e^-$ collisions in detail. 
We benchmark the performance of a number of sequential recombination 
algorithms: the classical $e^+e^-$ $k_\mathrm{t}$ algorithm~\cite{Catani:1991hj} 
used by the LEP experiments and SLD, the longitudinally invariant $k_{\mathrm{t}}$
algorithm~\cite{Catani:1993hr,Ellis:1993tq} used at hadron colliders 
and a generalization of the Valencia algorithm~\cite{Boronat:2014hva}, 
a robust $e^+e^-$ algorithm.

Several aspects of jet reconstruction performance are studied in
simulated events. In a particle-level study we estimate the size of 
perturbative and non-perturbative corrections to the jet energy. We establish
their dependence on the process and the centre-of-mass energy, and on the 
parameters of the jet algorithms. 
These particle-level simulations are also used to study the impact of 
energy deposited on the signal event by background processes. 
We conclude by using a realistic simulation 
of the CLIC detector and particle flow reconstruction to study the jet 
reconstruction performance in top quark pair production
and di-Higgs boson production.

The layout of this paper is as follows.
In Section~\ref{sec:challenges} we review the
challenges of jet reconstruction at high-energy lepton
colliders. In Section~\ref{sec:algorithms} the jet reconstruction algorithms
are introduced.
In Section~\ref{sec:corrections} the perturbative and non-perturbative 
corrections are studied. In Section~\ref{sec:toyexamples} we study the
response of the algorithms to the signal jet and the background.
Section~\ref{sec:simulation} presents the results of a full-simulation
study of a few benchmark processes.
In Section~\ref{sec:discussion} we discuss possible directions for future 
work and in Section~\ref{sec:conclusions} the most
important findings of this work are summarized.

\section{Challenges for jet reconstruction at high-energy $e^+e^-$ colliders}
\label{sec:challenges}

The experimental environment at previous lepton colliders, such as LEP and 
SLC, was very benign compared to that at hadron colliders. 
While this remains true at future high-energy lepton machines, 
jet reconstruction faces a number of new challenges.

\subsection{Multi-jet final states}

Future lepton colliders offer the possibility to study 
$2 \rightarrow 4$, $2 \rightarrow 6$, and even $ 2 \rightarrow 8$ 
processes. The dominant branching ratios of the $W$, $Z$ and Higgs
bosons are from hadronic decays. Final states with four 
jets, most notably $e^+ e^- \rightarrow Z h$, 
play a key role in the physics programme of any future 
electron-positron collider 
At high energy final states with six jets 
(e.g. $e^+ e^- \rightarrow t\bar{t} $), or
even eight or ten jets  (e.g. $e^+ e^- \rightarrow t\bar{t} h$), 
become important.
Imperfect clustering of final-state particles can affect the reconstruction 
of hadronic decays in an important way. 

The impact of incorrect assignments of final-state particles to jets is 
illustrated with an example. We consider the Higgsstrahlung process with
hadronic decays of $Z$- and Higgs bosons, where reliable reconstruction of 
$Z$ and $h$-boson candidates is the key to a precise measurement of
the Higgs boson couplings~\cite{Thomson:2015jda}. 
The hard scattering process $e^+ e^- \rightarrow Z h$ is simulated with 
the MadGraph\_aMC@NLO~\cite{Alwall:2014hca} package  
and the $Z \rightarrow q \bar{q}$ and 
$h \rightarrow b \bar{b}$ decays and subsequent hadronization with 
Pythia8~\cite{Sjostrand:2007gs}. No beam energy spread, 
initial-state-radiation, or modelling of background or the detector 
are included. Stable particles are clustered into
jets with the Durham algorithm (exclusive clustering with $N=$ 4). 

\begin{figure}[htbp!]
  {\centering 
\includegraphics[width=0.9\textwidth]{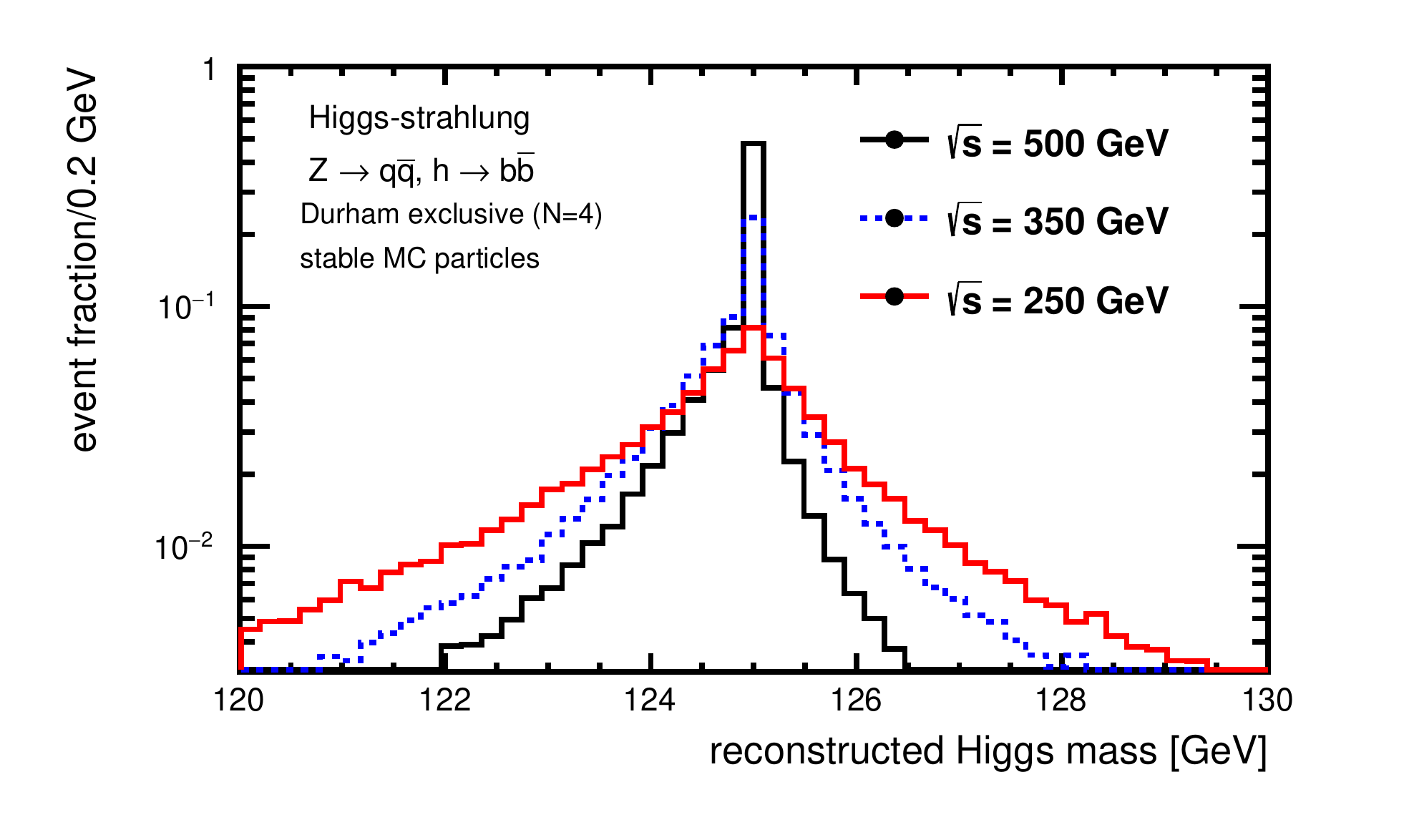}
\caption{The reconstructed Higgs boson mass peak in  $e^+ e^- \rightarrow Z h$,
 $Z \rightarrow q \bar{q}$, $h \rightarrow b \bar{b}$ events.
The three histograms correspond to centre-of-mass energies of 250~\gev{} 
(red, continuous line), 350~\gev{} (blue, dashed line) and 500~\gev{} 
(black, continuous line).}
\label{fig:higgsmasspeak}
}
\end{figure}

In Figure~\ref{fig:higgsmasspeak} the invariant mass distribution 
of the Higgs boson candidates is shown for three centre-of-mass energies.
We find that the distribution has a non-zero width, even in this relatively
perfect simulation. 
The finite resolution is purely due to imperfect clustering of
final-state particles into jets. The effect of confusion in jet 
clustering is less pronounced at higher centre-of-mass energy, 
as the greater boost of the $Z$- and Higgs bosons leads to a cleaner 
separation of the jets.

The most important challenge stems from the larger jet multiplicity.
In events with only two jets (i.e.\ $e^+ e^- \rightarrow Z h$ events
with $Z \rightarrow \nu \bar{\nu}$ and $h \rightarrow b \bar{b}$)
the clustering contribution to the mass resolution is negligible.
In the ILC and CLIC analyses of di-Higgs boson~\cite{LC-REP-2013-021} 
and $t\bar{t}h$ production~\cite{Price:2014oca,Redford:1690648} jet 
clustering is 
found to be the limiting factor for the Higgs mass resolution.

Finally, we note that as the centre-of-mass energy increases, 
$t$-channel processes become more important. The most obvious example is 
vector-boson fusion production of the Higgs boson.
The final-state products at high energy are strongly 
forward-peaked~\cite{Fuster:2009em} and special care is needed to ensure
robust jet reconstruction performance over the full polar angle
coverage of the experiment. Di-Higgs boson production through 
vector boson fusion ($e^+ e^- \rightarrow \nu \bar{\nu} hh$) presents
a double challenge of high jet multiplicity and forward jets.   
We therefore take this analysis at $\sqrt{s} = $ 3~\tev{} as a benchmark.


\subsection{Jet substructure}

The production of very energetic gauge bosons, Higgs bosons and top quarks 
with hadronic decays, collectively denoted as boosted objects, 
poses a challenge to the experiments at a future high-energy collider.
Whenever the energy of the decaying 
object exceeds its mass $m$ significantly
the highly collimated decay products typically 
cannot be resolved. In such cases, an analysis of the internal structure of jets
at a scale\footnote{At hadron colliders the scale $R$ is usually 
expressed in terms of the $\Delta R$ distance between two objects, defined
as $\Delta R = \sqrt{(\Delta \phi)^2 + (\Delta \eta)^2}$,
where $\phi$ and $\eta$ are the azimuthal angle and pseudo-rapidity} 
$R < 2 m/\pt$ is performed to identify the boosted 
object~\cite{Abdesselam:2010pt,Seymour:1993mx}.

At high-energy linear colliders, the separation of boosted $W$- 
and $Z$-bosons and the reconstruction of boosted top quarks challenge
the detector and reconstruction algorithms. The highly granular 
calorimeters~\cite{collaboration:2010rq,Adloff:2012gv} 
of the Linear Collider detector concepts and the particle flow paradigm 
are eminently suited for substructure analyses. We analyze the 
large-$R$ jet mass resolution in top quark pair production at 
$\sqrt{s} =$ 3~\tev{}, as a first exploration of the jet substructure 
performance of experiments at future lepton colliders.

\subsection{Beam-induced background}

While the environment at lepton colliders remains much more benign than
the pile-up conditions of high-energy hadron colliders, several background 
sources cannot be ignored in the detector design and 
evaluation of the performance of the linear collider experiments.  
The most relevant background source for jet reconstruction at 
linear $e^+e^-$ colliders is $\gamma \gamma \rightarrow $ {\em hadrons}
production~\cite{Linssen:2012hp}: photons emitted from the incoming 
electron and positron beams (bremsstrahlung and beamstrahlung) 
collide and produce {\em mini-jets} of hadrons\footnote{Other sources, 
in particular incoherent $e^+e^-$ pair production due to beamstrahlung photons, 
have a non-negligible impact on the design of the innermost detector elements, 
but can be ignored in the study of the jet reconstruction performance. A detailed discussion is found in Reference~\cite{Schulte:1999gg}.}. 
In high-energy colliders the probability to produce a mini-jet event
in a given bunch crossing is of the order of one. 

To evaluate the detector
performance, ILC and CLIC detector concepts superpose a number of 
$\gamma \gamma \rightarrow $ {\em hadrons} background events on 
the hard scattering process.
The distribution of the particles formed in $\gamma \gamma$ collisions
is forward-peaked, with approximately constant density per unit of rapidity
over the instrumented region (a feature also present in the {\em pile-up}
due to {\em minimum-bias} events in proton-proton collisions).
For CLIC at 3~\tev{} approximately 90\% of the 
energy is deposited in the endcap calorimeters and only 10\% in the 
central (barrel) regions of the experiment~\cite{Linssen:2012hp}.

The impact of the background on the performance depends on the bunch structure
of the accelerator and the read-out speed of the detector systems. In 
particular for machines based on radio-frequency cavities operated at 
room temperature the bunch spacing can be very small; CLIC envisages a 
bunch spacing of 0.5 ns. Background processes deposit 19~\tev{} in the 
detectors during a complete {\em bunch train} of 312 consecutive bunch 
crossings at $\sqrt{s}=$ 3~\tev{}~\cite{Linssen:2012hp}.
A selection based on the time stamp and transverse momentum of 
the reconstructed particles reduces this background to approximately 100~\gev{} 
on each reconstructed physics event. The relatively large bunch spacing of 
approximately 500 ns at the ILC allows the detector to distinguish 
individual bunch crossings.
In our full-simulation study simulated $\gamma \gamma \rightarrow $ 
{\em hadrons} events are overlaid on the signal events.

\subsection{Initial State Radiation}

\begin{figure}[htbp!]
  {\centering 
\includegraphics[width=0.9\textwidth]{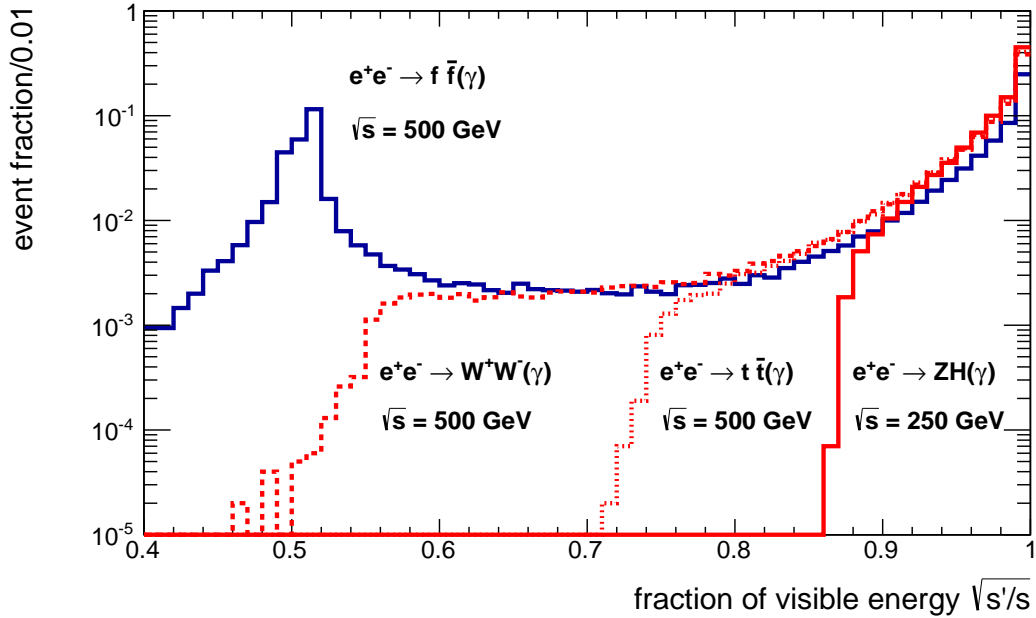}
\caption{The fraction of the visible energy (the energy carried by all final
state products except the photon) in several processes, where a photon radiated
off the initial or final-state particles may accompany the final state products. The distributions correspond
to pair production of light fermions
in association with a photon ($e^+e^- \rightarrow f \bar{f} (\gamma)$), 
$W$-boson and top-quark pair production ($e^+e^- \rightarrow W^+W^-(\gamma)$, 
$e^+e^- \rightarrow t \bar{t} (\gamma)$) and the Higgsstrahlung
process $e^+e^- \rightarrow Zh (\gamma)$. The centre-of-mass energy 
is indicated on the figure for each process.}
\label{fig:isr}
}
\end{figure}

In $e^+e^-$ annihilation processes the system formed by 
the final state products is, to first approximation,  produced at 
rest in the laboratory. Initial state radiation (ISR) photons emitted by the 
incoming electrons and positrons changes this picture somewhat. 
To estimate the magnitude of the boost introduced by ISR
we generate events using a parton-level calculation\footnote{The effects
of beam energy spread and beamstrahlung are not included.} in 
MadGraph\_aMC@NLO~\cite{Alwall:2014hca}. For each 2 $\rightarrow$ 2 
process $e^+e^- \rightarrow XY$ we include also the 2 $\rightarrow$ 3
process $e^+e^- \rightarrow XY \gamma$. In Fig.~\ref{fig:isr} the 
fraction of the energy carried by the $XY$ system is
shown for several 2 $\rightarrow$ 2 processes and 
for several centre-of-mass energies.
For most $s$-channel processes, the ISR photon energy spectrum
falls off very rapidly and the visible energy distribution displays 
a sharp peak at 1.

The boost of the system along the $z$-axis due to ISR remains 
relatively small close to the production threshold:
for $e^+e^- \rightarrow Zh (\gamma)$ at 
$\sqrt{s} =$ 250~\gev{} and $e^+e^- \rightarrow t \bar{t} (\gamma)$ 
at 500~\gev{}  $\beta_z = v_z/c$ is smaller
than 0.1 in over 95\% and 90\% of the events, respectively\footnote{Compared 
to hadron colliders this boost is very small indeed: 
for di-jet production at the LHC $\beta_z $ of the 
di-jet system is very close to 1 and even a massive system such as a top 
quark pair acquires a typical $\beta_z =$ 0.5.}. 
For processes with a cross section that grows with $\sqrt{s}$ 
the peak is even narrower. The role of ISR is only significant for radiative 
return to the Z in the process $e^+e^- \rightarrow f \bar{f} (\gamma)$, 
where $f$ is any fermion with mass less than half that of the $Z$ boson.

At linear colliders with very narrow beams the luminosity spectrum 
displays a sizeable tail towards lower centre-of-mass 
energy~\cite{Poss:2013oea}. Beam energy spread and beamstrahlung 
may therefore lead to a pronounced boost of the visible final state 
objects in a small fraction of events. This effect is 
included in the full simulation study of Section~\ref{sec:simulation}.

\section{Jet reconstruction algorithms}
\label{sec:algorithms}

In this section the jet algorithms considered in this paper are
introduced. We discuss the most important differences and their
implications for the performance.
Three classes of sequential recombination algorithms are considered here:
\begin{itemize}
\item The classical $e^+e^-$ algorithms~\cite{Catani:1991hj} and their generalization with beam jets~\cite{Cacciari:2011ma}
\item The longitudinally invariant algorithms developed for hadron colliders~\cite{Catani:1993hr,Ellis:1993tq}
\item The Valencia algorithm proposed in a previous publication~\cite{Boronat:2014hva}, which is further generalized here.
\end{itemize}

\subsection{The VLC algorithm}
\label{sec:vlc}

In Ref.~\cite{Boronat:2014hva} a robust jet reconstruction algorithm
was proposed, that maintains a Durham-like distance criterion based on 
energy and polar angle. It achieves a background resilience that can
compete with the longitudinally invariant $k_t$ algorithm. 
Here, we further generalize the definition of the algorithm.

The VLC algorithm has the following inter-particle distance:
\begin{equation} 
 d_{ij} = 2 \min{(E_i^{2 \beta},E_j^{2 \beta})} (1 - \cos{\theta_{ij}})/R^2,
\label{eq:distance_vlc}
\end{equation}
where $R$ is the radius or resolution parameter.
For $\beta = $1 the distance is given by the transverse momentum 
squared of the softer of the two particles relative to the harder one, 
as in the Durham algorithm. 

The beam distance of the algorithm is:
\begin{equation}
 d_{i\mathrm{B}} = E_i^{2\beta} \sin^{2\gamma}{\theta_{i\mathrm{B}}}, 
\end{equation}
where $\theta_{i\mathrm{B}}$ is the angle with respect to the beam axis, i.e.\ the 
polar angle. 

The two parameters $\beta$ and $\gamma$ allow independent control
of the clustering order and the background resilience\footnote{The first
version of the algorithm~\cite{Boronat:2014hva}
had a single parameter $\beta$. Equation~\ref{eq:distance_vlc} furthermore 
differs by a factor two from the inter-particle distance of 
Ref.~\cite{Boronat:2014hva}. To distinguish the two algorithms we refer to the 
more general expression as the VLC algorithm, while the name Valencia is 
reserved for the setting $\beta = \gamma$ that recovers the first
proposal (adjusting $R$ by factor $\sqrt{2}$).}. 
The $\beta$
and $\gamma$ parameters are real numbers that can take any value. 
For $\beta=\gamma=1$ the expression for the beam distance simplifies
to $d_{i\mathrm{B}} = E^2 \sin^2{\theta_{i\mathrm{B}}} = p^2_{\mathrm{T}i}$. We discuss the
impact of different choices in Sec~\ref{sec:betagamma}.

This new version of the algorithm fulfils the standard 
IR-safety tests of the FastJet team.
The VLC algorithm is available as a plug-in for the
FastJet~\cite{Cacciari:2005hq,Cacciari:2011ma} package. 
The code can be obtained from the ``contrib'' area~\cite{fjcontrib}.

\subsection{Comparison of the distance criteria}
\label{sec:comparison}

The generalized distance criteria for three families of algorithms are summarized in 
Tab.~\ref{tab:distance_criteria}. 
\begin{table}
\caption{Summary of the distance criteria used in sequential recombination algorithms. Generalized 
inter-particle and beam distances are given for three main classes of algorithms: the classical $e^+e^-$ 
algorithms (comprising a version with beam jets of the Cambridge~\cite{Dokshitzer:1997in} and Durham 
algorithms), the longitudinally invariant algorithms used at hadron colliders, which comprise longitudinally 
invariant $k_t$, Cambridge-Aachen and anti-$k_t$ and the robust $e^+e^-$ algorithms introduced in 
Section~\ref{sec:vlc}. \label{tab:distance_criteria}}
\begin{tabular}{lccc}
algorithm & generalized $e^+e^-$ & longitudinally  invariant &   robust  $e^+e^-$      \\ \hline
                    &              &                     &           \\
distance $d_{ij}$    &   
       $ 2 \min (E_i^{2n}, E_j^{2n}) \frac{1 - \cos \theta_{ij}}{1 - \cos R} $  &
      $  \min (p_{\mathrm{T},i}^{2n}, p_{\mathrm{T},j}^{2n}) \frac{\Delta R_{ij}^{2}}{R^{2}} $ &
      $  2 \min (E_i^{2\beta}, E_j^{2\beta}) \frac{1 - \cos \theta_{ij}}{R^{2}} $  \\
                        &          &                       &           \\
beam distance $d_{iB}$ & $ E_i^{2n}$ & $ p_{\mathrm{T},i}^{2n}$  & $E_i^{2\beta} \sin^{2\gamma}{\theta_{i\mathrm{B}}}$ \\
                      &            &                       &                  \\ \hline
\end{tabular}
\end{table}


For all algorithms the clustering order can be modified by an appropriate
choice of the $n$ in the exponent of the energy (or $p_{\mathrm{T}}$) in 
the inter-particle distance ($\beta$ in the VLC algorithm). 
The Durham (or $k_t$) algorithm, with $n=1$,
clusters pairs of particles starting with those that are soft and collinear 
(i.e.\ the inverse of the virtuality-ordered emission
during the parton shower). Choosing $n=$ 0 yields the Cambridge/Aachen 
algorithm, that has a purely angular distance criterion. The anti-$k_t$
algorithm has $n=-$1.

In the generalized algorithms the area of jets is limited. Any particle 
with a {\em beam distance}~\cite{Catani:1992zp} smaller than the 
distance to any other particle is associated with the beam jet, 
and therefore not considered part of the visible final state. 
The radius parameter $R$ governs the relative size of the
interparticle and beam distance and thus determines the size of
the jet. This modification renders
jet reconstruction very resilient to backgrounds.

The generalized $e^+e^-$ algorithm and the VLC algorithm have virtually
the same inter-particle distance. However, the radius parameter $R$ is
redefined: the inter-particle distance $d_{ij}$ denominator 
is $R^2$ instead of $ 1 - \cos{R}$. The hadron collider 
algorithms replace the 
particle energy $E_i$ and angle $\theta_{ij}$ with quantities 
that are invariant under
boosts along the beam axis, the transverse 
momentum $p_{\mathrm{T}i}$ and the distance $\Delta R_{ij} =  \sqrt{(\Delta \phi)^2 + (\Delta y)^2}$, where $\phi$ is the azimuthal angle in the usual cylindrical coordinates and $y$ denotes the rapidity.

\begin{figure*}[h!]
  {\centering 
\includegraphics[width=\textwidth]{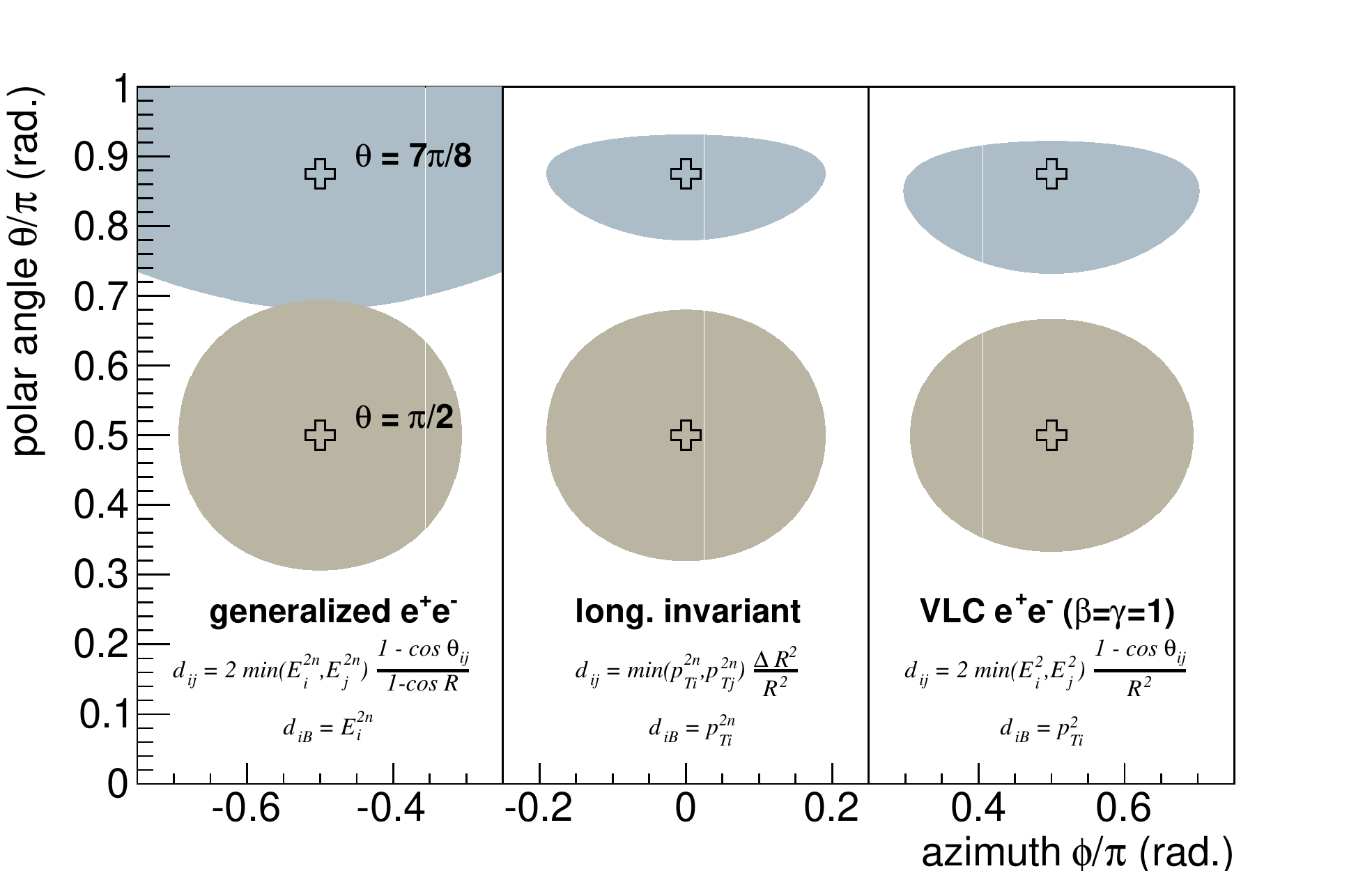}
\caption{The area or {\em footprint} of jets reconstructed with $R=$ 0.5 with the three major families of sequential recombination algorithms. The two shaded areas in each column correspond to a jet in the central detector ($\theta = \pi/2$) and to a forward jet ($\theta = 7\pi/8$). The jet axis is indicated with a cross. }
	\label{fig:jet_sizes}
	}
	\end{figure*}

Detailed studies~\cite{Linssen:2012hp,Marshall:2012ry} show that the
the longitudinally invariant $k_t$ algorithm is much more resilient to the 
$\gamma \gamma \rightarrow$ {\em hadrons} background than the 
classical and generalized $e^+e^-$ algorithms.

For each of the algorithms the 
catchment areas of a single central and forward jet with $n=$1 and $R=$ 0.5 
are indicated in Figure~\ref{fig:jet_sizes}. The footprint of the central jet 
(at $\theta=\pi/2$) is approximately circular for all algorithms. The area
of the jet in the forward detector (at $\theta=7\pi/8$) shrinks considerably
for the longitudinally invariant algorithms and the VLC algorithm. 
The reduced exposure in this region, where backgrounds are most pronounced, is
the crucial feature for the enhanced resilience of these 
algorithms.

An analytical understanding of this property can
be obtained by considering two test particles with energies $E_i$ and $E_j$ 
and separated by a fixed angle $\Omega_{ij}$. For the generalized $e^+e^-$ algorithms, 
both the distance $d_{ij}$ between the two particles and the
ratio $d_{ij}/d_{i\mathrm{B}}$ of the inter-particle distance and the beam distance 
are independent of polar and azimuthal angle. 
For the longitudinally invariant algorithms 
the ratio $d_{ij}/d_{i\mathrm{B}}$ increases (while the inter-particle distance decreases) 
as the two-particle system is rotated into the forward region. 
Finally, for the VLC algorithm the ratio $d_{ij}/d_{i\mathrm{B}}$ increases
as $1/\sin^{2\gamma} \theta$ in the forward region, with a slope that 
is similar to that of the longitudinally invariant algorithms for $\gamma =$ 1. 
The distance $d_{ij}$ is constant, as in classical $e^+e^-$ algorithms.

A closer comparison of the shape of the footprint of the longitudinally 
invariant algorithms and the VLC algorithm show that, given identical
jet axes, the former extend further into the forward region. 
This causes a slight difference in background resilience 
of both classes of algorithms. 

\subsection{Interpolation between algorithms}
\label{sec:betagamma}

The two parameters $\beta$ and $\gamma$ of the VLC algorithm 
allow one to tailor the algorithm to a specific application. As these
parameters are real numbers, one can interpolate smoothly between 
different clustering schemes.

The $\beta$-parameter that exponentiates the energy in inter-particle and
beam distance governs the clustering order (similar to the exponent $n$
in the generalized $k_t$ algorithm). For $\beta = $1 clustering
starts with soft, collinear radiation. Choosing $\beta=$0 yields purely 
angular clustering, while $\beta=-$1 corresponds to clustering starting 
from hard, collinear radiation. These integer choices of $\beta$ correspond
to $k_t$, Cambridge/Aachen and anti-$k_t$ clustering, respectively. 
Non-integer values of $\beta$ interpolate smoothly between these three schemes.

The parameter $\gamma$ in the exponent of the beam distance of 
the VLC algorithm provides a handle to control the shrinking of the
jet catchment area in the forward regions of the experiment.
After setting the $R$-parameter to the optimal value for central
jets, the area of forward jets can be {\em tuned} by the choice
of $\gamma$ to ensure the required background resilience.

\begin{figure}[htbp!]
{\centering 
\includegraphics[width=0.5\textwidth]{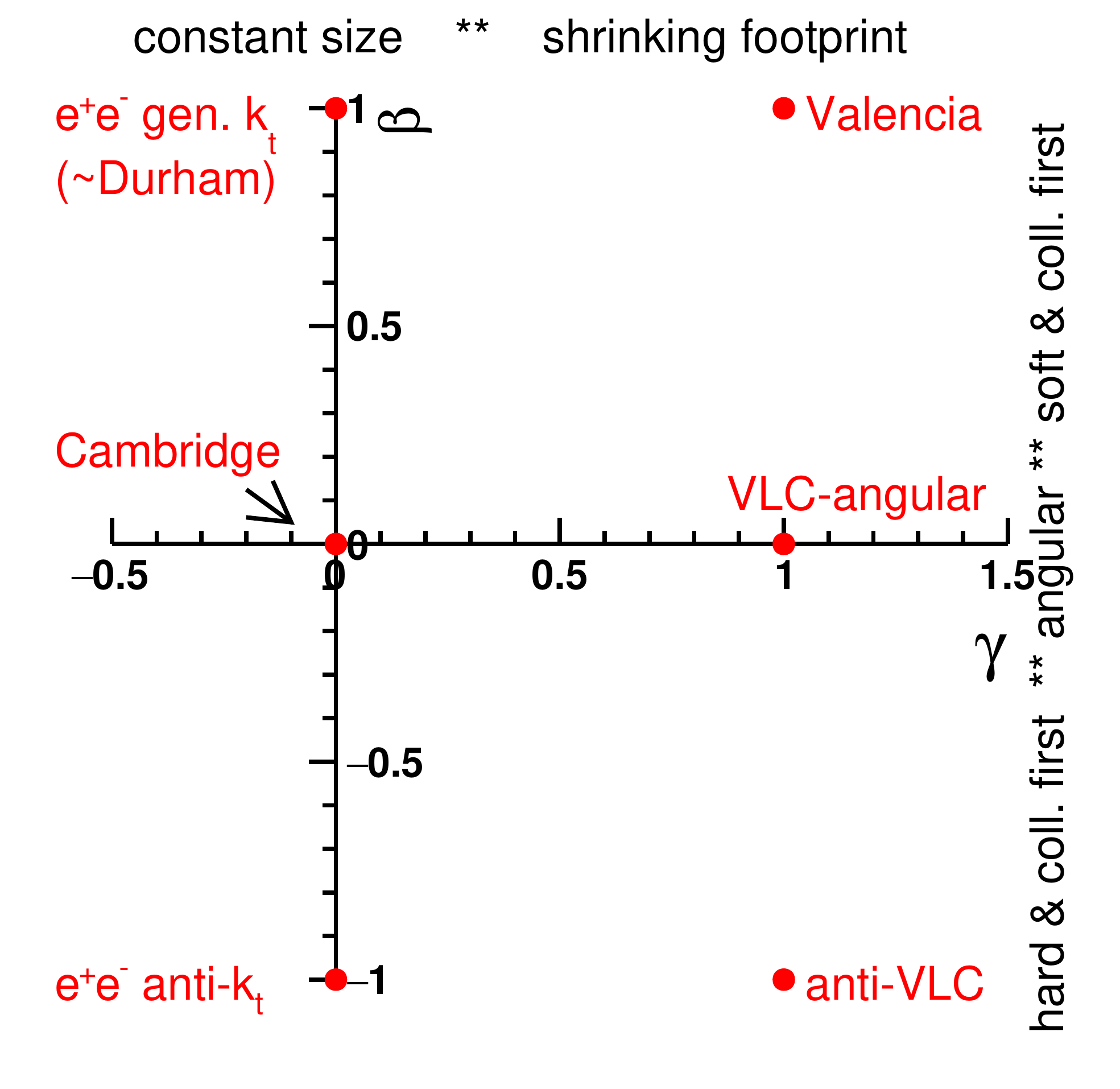}
\caption{Diagram of the parameter space spanned by exponents $\beta$ and 
$\gamma$ of the VLC algorithm. On the y-axis generalizations with 
beam jets of the LEP/SLD algorithms are found, with the Cambridge 
algorithm with angular ordering at the 
origin and the Durham or $k_t$ algorithm at $\beta=1$. 
Choosing $\beta=$ -1 yields reverses the clustering order 
(like in anti-$k_{\mathrm{t}}$ algorithm~\cite{Cacciari:2008gp}). 
Choosing non-zero and positive values for $\gamma$ yields robust algorithms
with a shrinking jet area in the forward region. }
	\label{fig:diagram}
}
\end{figure}

We have seen that $\gamma = $ 1 yields forward jets with a similar size
of those of the longitudinally invariant algorithms for hadron 
colliders\footnote{In both algorithms the ratio $d_{ij}/d_{i\mathrm{B}}$ 
for two test particles separated by a constant angle depends on 
the polar angle of the system. In the VLC algorithm the ratio is 
proportional to $\sin^{-2\gamma} \theta$. In longitudinally invariant 
algorithms the ratio follows the evolution of the pseudo-rapidity
and grows approximately as $\eta^{2} = \log^{2} \tan{\theta/2}$
in the forward region. Both gives rise to qualitative the same behaviour.}.
Values of $\gamma$ greater than 1 further enhance the rise
of the $\frac{d_{ij}}{d_{i\mathrm{B}}}$ ratio in the forward region, causing 
the jet footprint to shrink faster. Values between 0 and 1 yield a slower
decrease of the area when the polar angle goes to 0 or $\pi$. 

For $\gamma=$ 0, $d_{i\mathrm{B}} = E_{i}^{2\beta}$ and we retrieve
the generalized $e^+e^-$ algorithms with constant angular opening:
the generalized Cambridge algorithm~\cite{Dokshitzer:1997in} for $\beta=$ 0
and generalized $k_{\mathrm{t}}$ or Durham~\cite{Catani:1991hj} 
for $\beta=$ 1. Choosing $\beta=$ -1
yields an $e^+e^-$ variant of the anti-$k_{\mathrm{t}}$ 
algorithm~\cite{Cacciari:2008gp}.
A schematic overview of the algorithms in $(\beta, \gamma)$ space is given
in Figure~\ref{fig:diagram}.

\section{Jet energy corrections}
\label{sec:corrections}

Before we turn to a detailed simulation including overlaid backgrounds 
and a model for the detector response, we study the perturbative 
and non-perturbative jet energy corrections of the algorithms. 
Both types of corrections 
are closely connected to the jet area~\cite{Dasgupta:2007wa}. 
In this Section we quantify their impact, following the analysis of 
Ref.~\cite{Dasgupta:2007wa}. This first exploration of the stability
of the algorithms should be extended in future work to quantify the
impact of next-to-leading correction, as performed for instance 
in Ref.~\cite{Bethke:1991wk}. Also the robustness of the conclusions
for a variety of different sets of parameters ({\em tunes}) 
of the Monte Carlo simulation merits further study.

\subsection{Monte Carlo setup}
\label{sec:mcparticle}

The Monte Carlo simulation chain uses the MadGraph5\_aMC@NLO 
package~\cite{Alwall:2014hca} to generate the matrix elements of the hard 
scattering $2 \rightarrow 2$ event. Several processes are studied, but
results in this Section focus on $e^+e^- \rightarrow q\bar{q}$ at 
$\sqrt{s}=$ 250~\gev{} and $e^+e^- \rightarrow t\bar{t}$ with 
fully hadronic top decays at $\sqrt{s}=$ 3~\tev.
The four-vectors of the outgoing quarks are fed into 
Pythia 8.180~\cite{Sjostrand:2007gs}, with the default tune to LEP data,
that performs the simulation of top quark and $W$ boson decays, the 
parton shower and hadronization. No detector simulation is performed 
and initial state
radiation and beam energy spread are not included in the simulation. 
Particles or partons from the Pythia event record are clustered using 
FastJet 3.0.6~\cite{Cacciari:2011ma} exclusive clustering with $N=$ 2.
The default (``E-scheme'') recombination algorithm is used to
merge (pseudo-) jets.

\subsection{Definition of response and resolution}
\label{sec:definitions}

The jet energy and mass distributions often display substantial non-Gaussian
tails and the choice of robust estimators has non-trivial implications. 
To estimate the centre of the distribution (i.e.\ the response) 
the mean and median are used. In some cases we present both, to 
give an indication for the skewness of the distribution. The width of the 
distribution (resolution) is estimated using the inter-quantile range 
$\mathrm{IQR}_{34}$, that measures half the width of the interval centered 
on the median that contains 68\% of all jets. 
We also use $\mathrm{RMS}_{90}$, the 
root-mean-square of the values after discarding 5\% outliers in both 
the low and high tails of the distribution.

\subsection{Perturbative corrections}

Following Reference~\cite{Dasgupta:2007wa} we estimate the
total energy correction by comparing the parton from the hard scatter 
to the jet of stable particles.
For jets of finite size this correction is dominated by energy that {\em leaks}
out of the jet. We indeed find that the distribution of the difference of 
parton and jet energy is asymmetric, with a long tail towards negative 
corrections, where the parton energy is larger than the energy captured 
in the jet. This energy leakage is most pronounced for jets with a 
small radius parameter, as expected.

\begin{figure}[htbp!] 
{\centering 
 \includegraphics[width=0.49\textwidth]{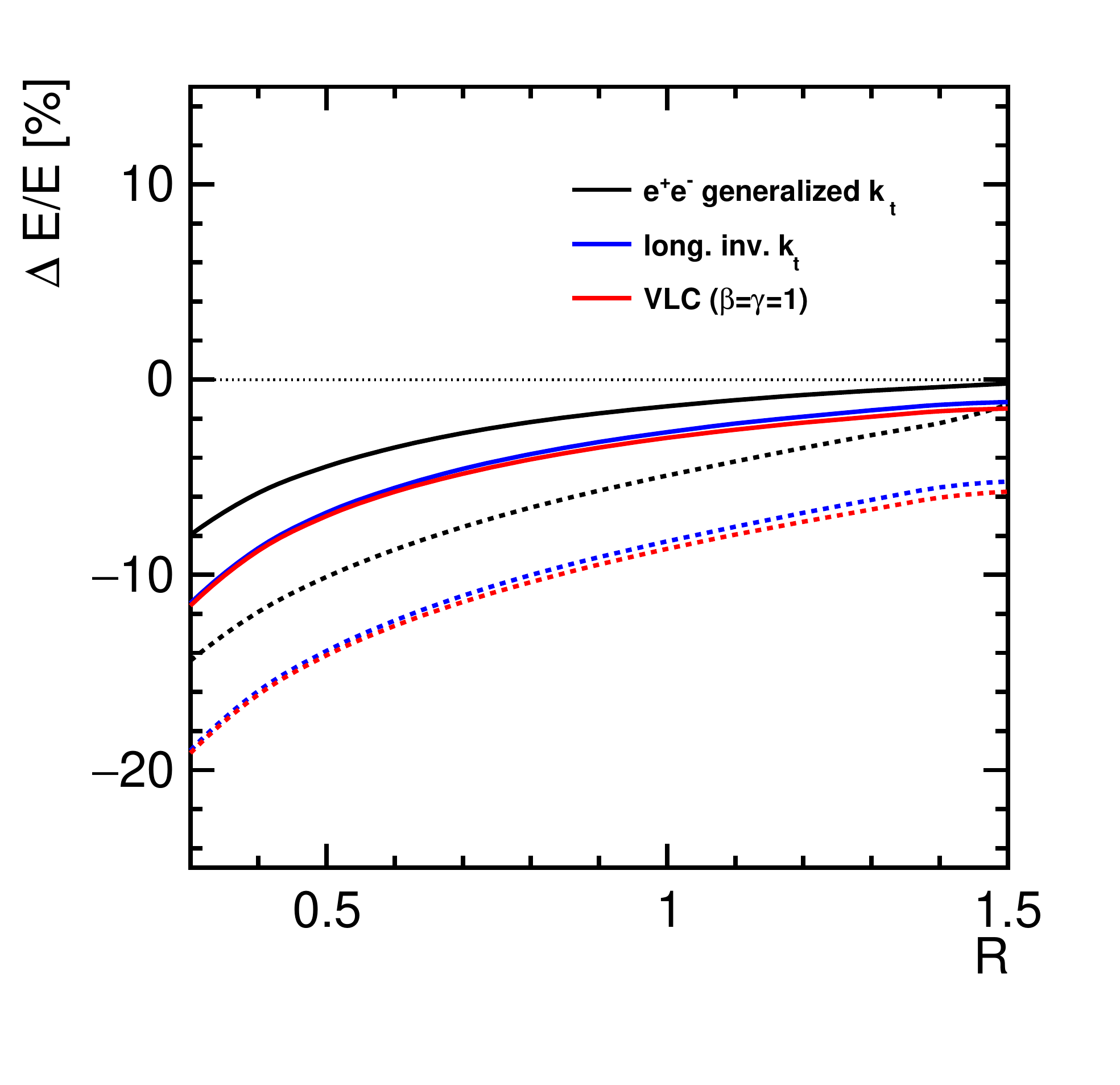} 
 \includegraphics[width=0.49\textwidth]{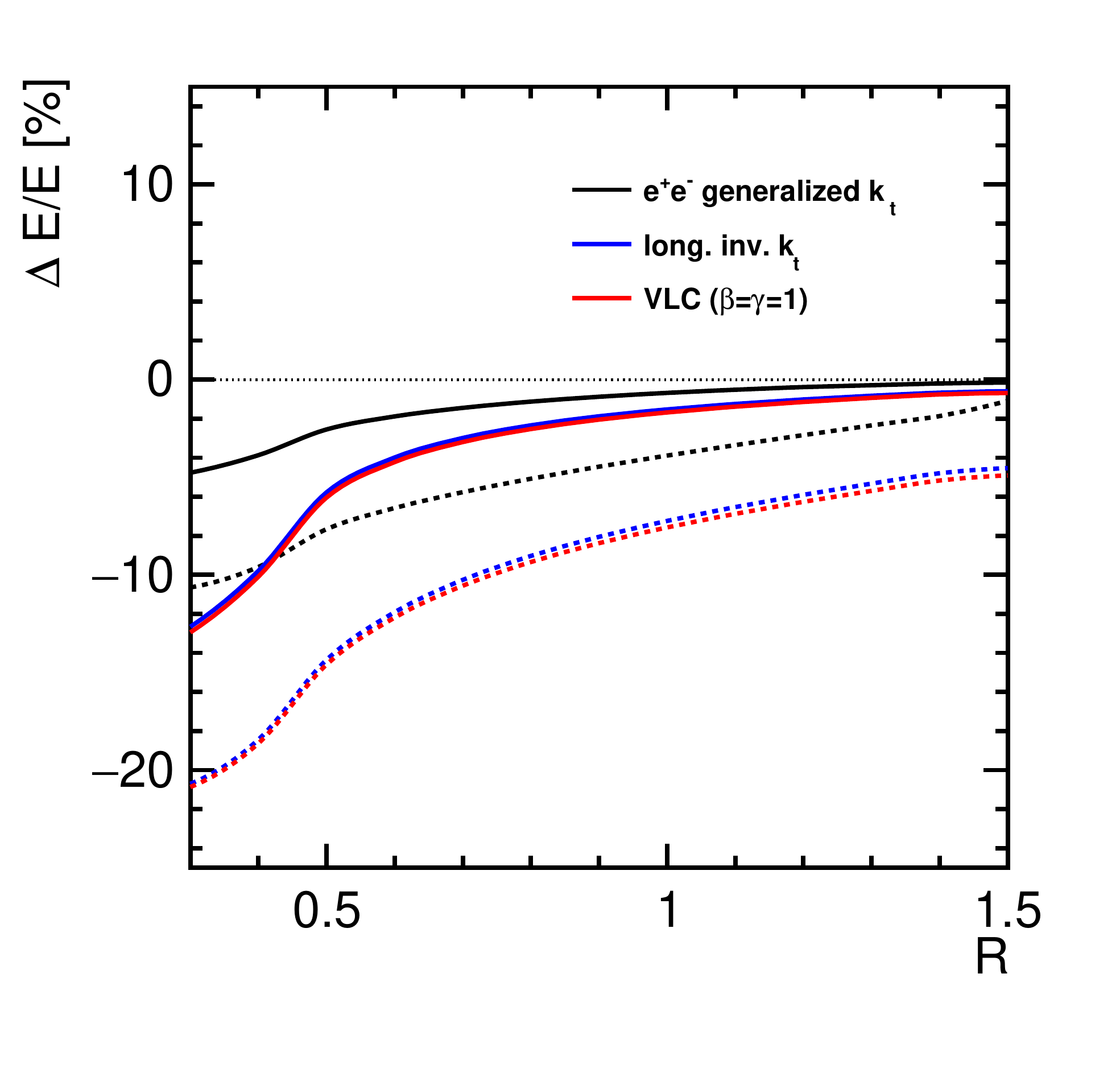} 
  \caption{Jet energy correction as a function of the 
jet radius parameter $R$ in $e^+ e^-\rightarrow q \bar{q}$ production at 
$\sqrt{s} =$ 250~\gev (left panel) and $e^+ e^-\rightarrow t \bar{t}$
production at $\sqrt{s}=$ 3~\tev (right panel). 
The continuous line corresponds to the median relative correction, the
dashed line to the mean. Results are shown for three algorithms: 
the generalized $e^+e^-$ algorithm, 
the longitudinally invariant $k_t$ algorithm and the VLC algorithm 
with $\beta= $ 1. The statistical uncertainties of the results 
are negligible and are not indicated.}
  \label{fig:energycorr}
}
\end{figure}

In Figure~\ref{fig:energycorr} the average (dashed line) and median 
(continuous line) relative energy correction are presented.
The left plot corresponds to $e^+ e^-\rightarrow q \bar{q}$ 
collisions at relatively low energy ($\sqrt{s}= $ 250~\gev), while
the right plot corresponds to  $e^+ e^-\rightarrow t \bar{t}$
 at $\sqrt{s}= $ 3~\tev. At a quantitative level the results show 
some dependence on the process, centre-of-mass energy and the generator 
tune for which they are obtained, but qualitatively the same pattern 
emerges in all cases. The energy correction decreases as the catchment
area of the jet increases. 

The energy corrections for the generalized $e^+ e^-$ algorithm 
vanish relatively rapidly, with the median correction reaching sub-\% level 
for R $\sim$ 1. The VLC and longitudinally invariant $k_t$ algorithm
show much slower convergence towards zero correction. This is entirely
due to jets close to the beam axis. For central jets the
three classes of algorithms yield identical results (within the statistical 
accuracy). The VLC and $k_t$ algorithms have similar footprints and, indeed,
very similar energy corrections.

The clustering
order (as controlled by $n$ in the generalized algorithm and by $\beta$ in
the VLC algorithm has a minor impact on the energy corrections.
The (inclusive) Cambridge/Aachen algorithm and anti-$k_t$ algorithm 
give similar results to the $k_t$ variants of the same algorithm shown here.

\subsection{Non-perturbative corrections}

\begin{figure}[htbp!] 
{\centering 
 \includegraphics[width=0.49\textwidth]{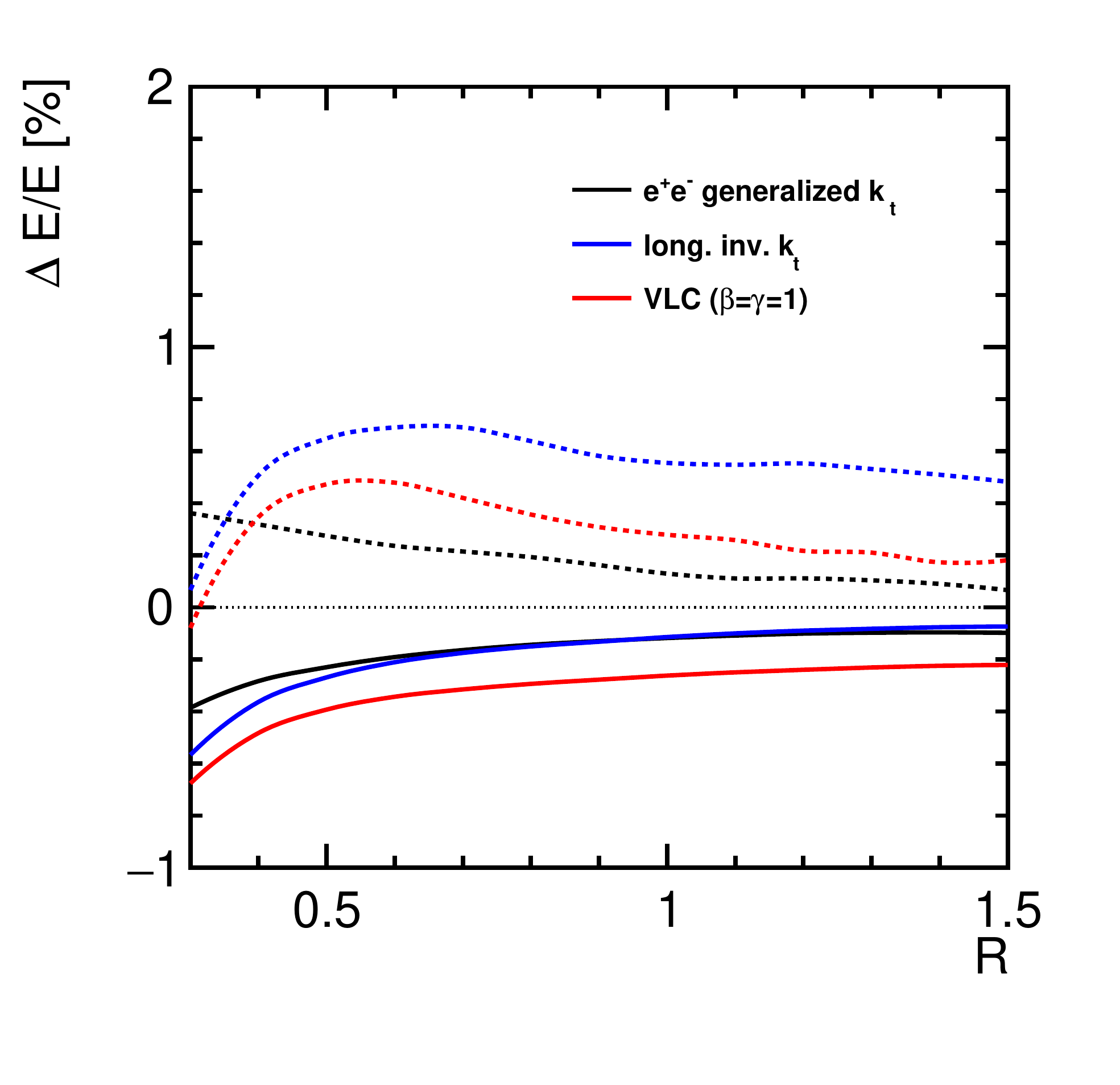}
 \includegraphics[width=0.49\textwidth]{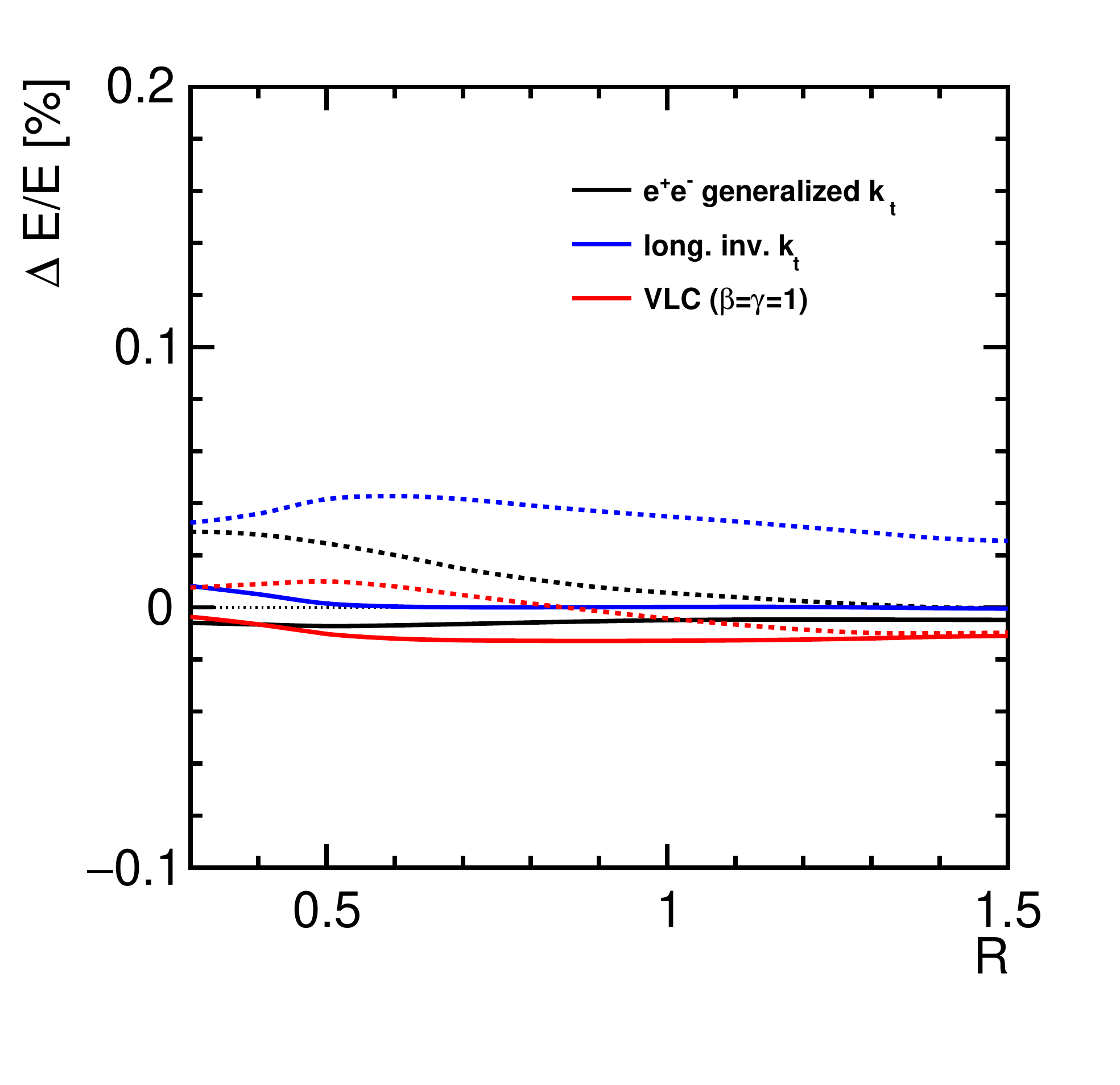}
  \caption{Non-perturbative jet energy corrections to the jet energy 
as a function of the 
jet radius parameter $R$ in $e^+ e^-\rightarrow q \bar{q}$ production at 
$\sqrt{s} =$ 250~\gev{} (left panel) and $e^+ e^-\rightarrow t \bar{t}$
production at $\sqrt{s}=$ 3~\tev{} (right panel). 
The continuous line corresponds to the median relative correction, the
dashed line to the mean. Results are shown for three algorithms: 
the generalized $e^+e^-$ algorithm, 
the longitudinally invariant $k_t$ algorithm and the VLC algorithm 
with $\beta= $ 1. The statistical uncertainties of the results 
are negligible and are not indicated.}
  \label{fig:energycorr_nonpert}
}
\end{figure}
 
The largest part of the jet energy correction due to the finite size is 
amenable to perturbative calculations. A small residual correction
is related to the hadronization and must be extracted from (or tuned to) data.
The non-perturbative energy correction is estimated as the difference
between the energy of the parton-level jet, clustering all partons
before hadronization, and the jet reconstructed from stable final-state 
particles. The difference in energy between the parton-level and
particle-level jet is typically small, but the distribution is offset
from 0 and has a long asymmetric tail. Mean and median are again
different and even have opposite signs.

The dependence of this correction on $R$ is shown in 
Figure~\ref{fig:energycorr_nonpert}. The non-perturbative part is
very small compared to the total correction. It is well below 1\%
at $\sqrt{s} = $ 250~\gev, for any value of $R$ studied here. 
For high-energy collisions the correction is well below the per mille
level. The generalized $e^+e^-$ algorithm again has the best convergence,
while for both VLC and longitudinally invariant $k_t$ the median 
or mean remain sizeable even for $R=$ 1.5. 

\subsection{Jet mass corrections} 

The previous discussion has focussed on the jet energy response. Corrections
to other jet properties may also be important. Here, we study the corrections
to the jet mass\footnote{The jet mass is defined as the invariant mass
formed by the vector-sum of the momenta of the (massless) jet constituents.}, 
which can be taken as a proxy for the substructure of the jet.
The non-perturbative jet mass correction is defined (analogously to 
the non-perturbative energy correction) as the difference between the 
masses of the parton-level and particle-level jet. 

\begin{figure}[htbp!] 
{\centering 
\includegraphics[width=0.49\textwidth]{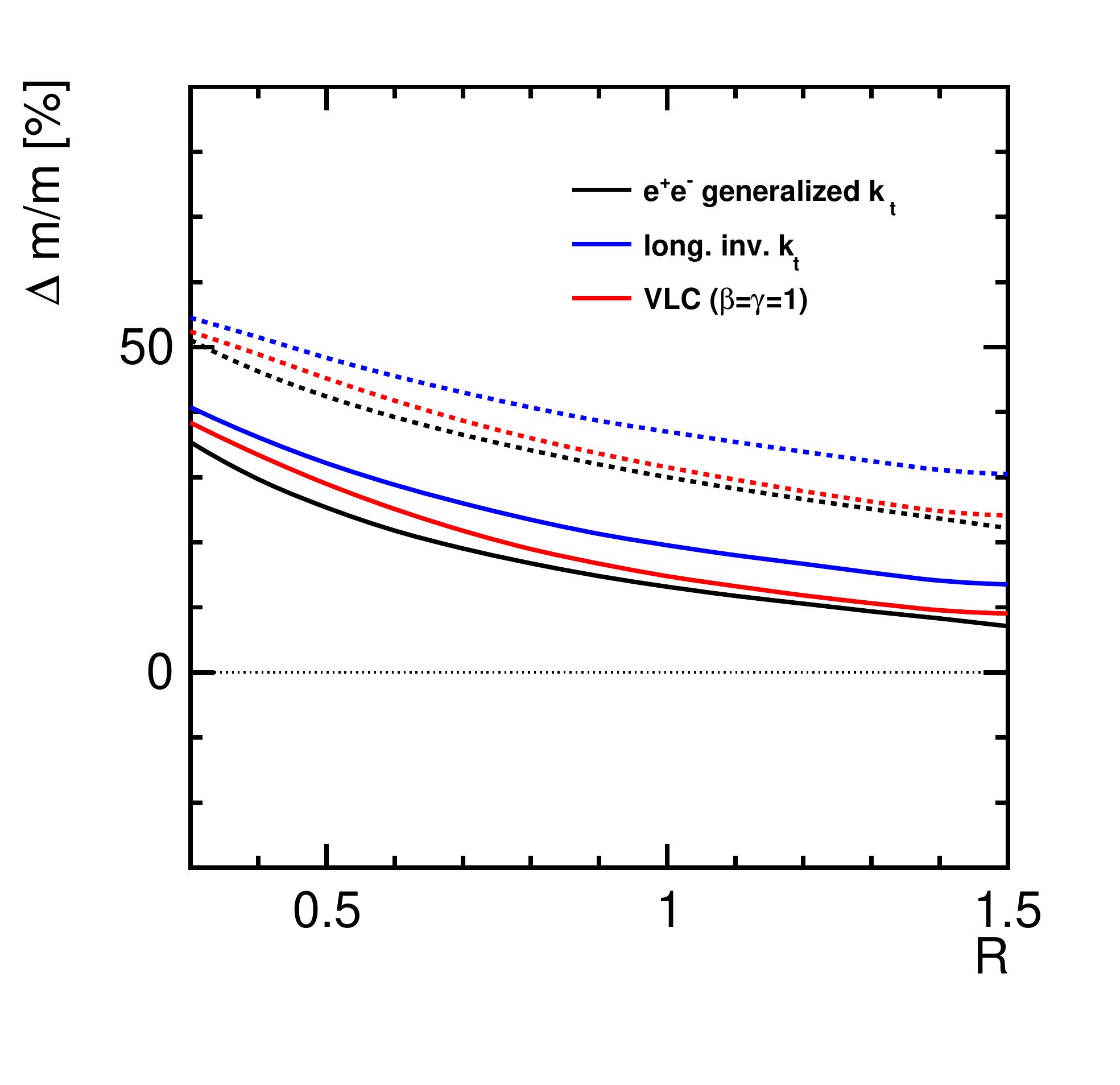} 
 \includegraphics[width=0.49\textwidth]{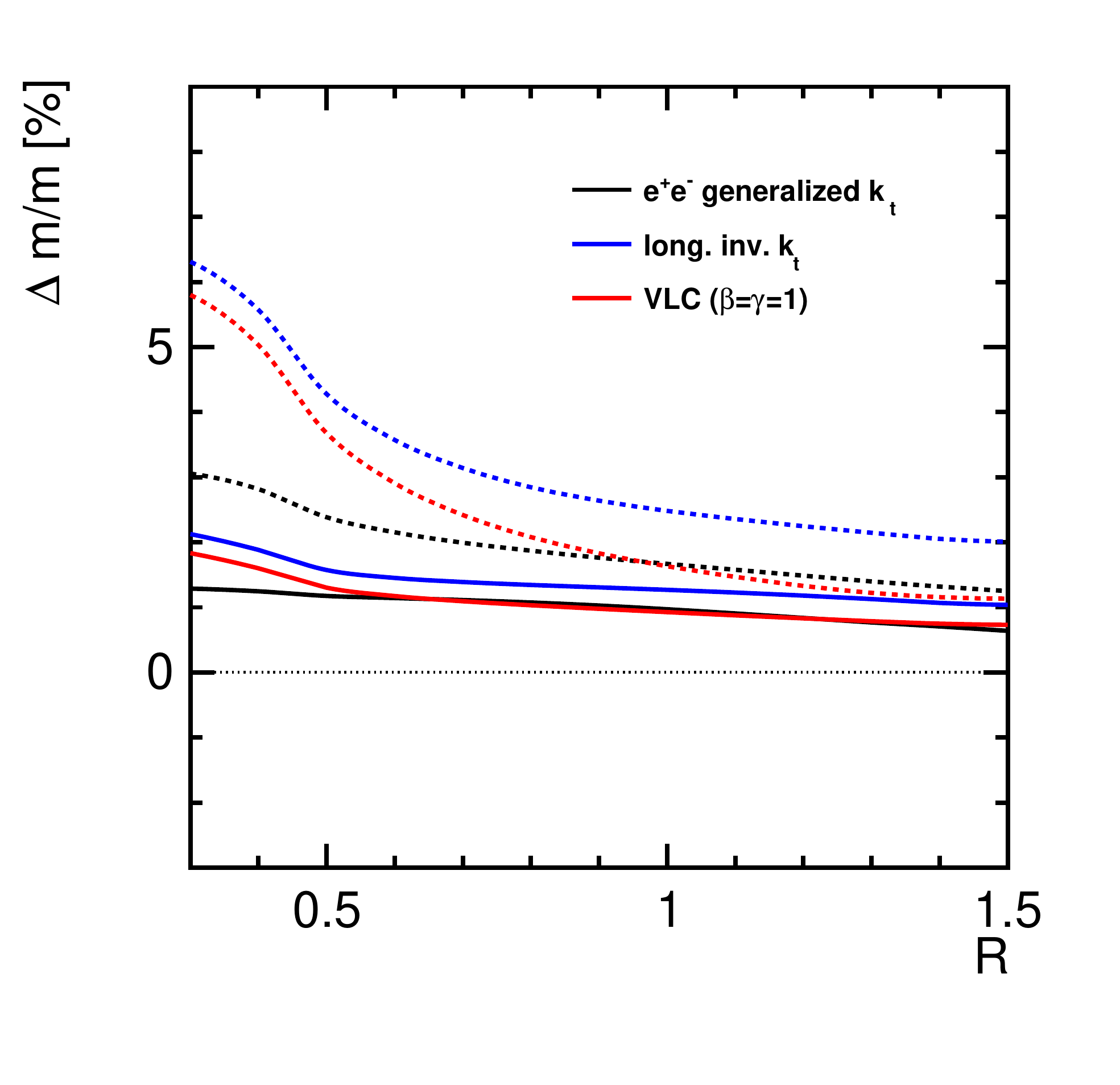}
  \caption{Non-perturbative corrections to the jet mass as a function of the 
jet radius parameter $R$ in $e^+ e^-\rightarrow q \bar{q}$ production at 
$\sqrt{s} =$ 250~\gev{} (left panel) and $e^+ e^-\rightarrow t \bar{t}$
production at $\sqrt{s}=$ 3~\tev{} (right panel). 
The continuous line corresponds to the median relative correction, the
dashed line to the mean. Results are shown for three algorithms: 
the generalized $e^+e^-$ algorithm, 
the longitudinally invariant $k_t$ algorithm and the VLC algorithm 
with $\beta= $ 1. The statistical uncertainties of the results 
are negligible and are not indicated.}
  \label{fig:energycorr_mass}
}
\end{figure}

The dependence on
the radius parameter $R$ is shown in Figure~\ref{fig:energycorr_mass}.
The non-perturbative contribution to the jet mass is quite large. The relative
correction can be several tens of \% at low energy up to $R=$ 1. It drops to 
a few \% for $\sqrt{s}=$ 3~\tev. In this case, the algorithms with the 
$e^+ e^-$ inter-particle distance (generalized Durham and VLC) converge 
slightly faster than the longitudinally invariant algorithms.

\section{Particle-level results}
\label{sec:toyexamples}

In this Section the response of several algorithms is studied on  
simulated $e^+ e^-\rightarrow t \bar{t}$ events at $\sqrt{s} = $ 3~\tev.
Clustering is exclusive, with $N=$ 2. Both highly boosted top quarks
are reconstructed as a single, large-$R$ jet. 
We gain insight in the impact of the background 
by superposing randomly distributed background on the signal events.
To this end the Monte Carlo setup described in Section~\ref{sec:mcparticle}
is extended with a simple mechanism to superpose a random energy flow
on the signal event. 

\subsection{Jet energy response without background}

Before we study the impact of the background, the response of the jet
algorithms to the signal event is estimated. The energy response
is determined as the median reconstructed energy. The mass
distribution has a sharp peak at the top quark mass and a long tail 
towards larger masses. The mass response is therefore estimated
as the mean reconstructed jet mass. In both cases the response
of the Durham algorithm -- which clusters all final state
particles into the jets -- is taken as a reference. 
The reconstructed energy is divided by 1.5~\tev, the 
reconstructed jet mass by the average jet mass of $\sim$ 370~\gev. 

\begin{figure}[htbp!]
{\centering 
\includegraphics[width=0.49\textwidth]{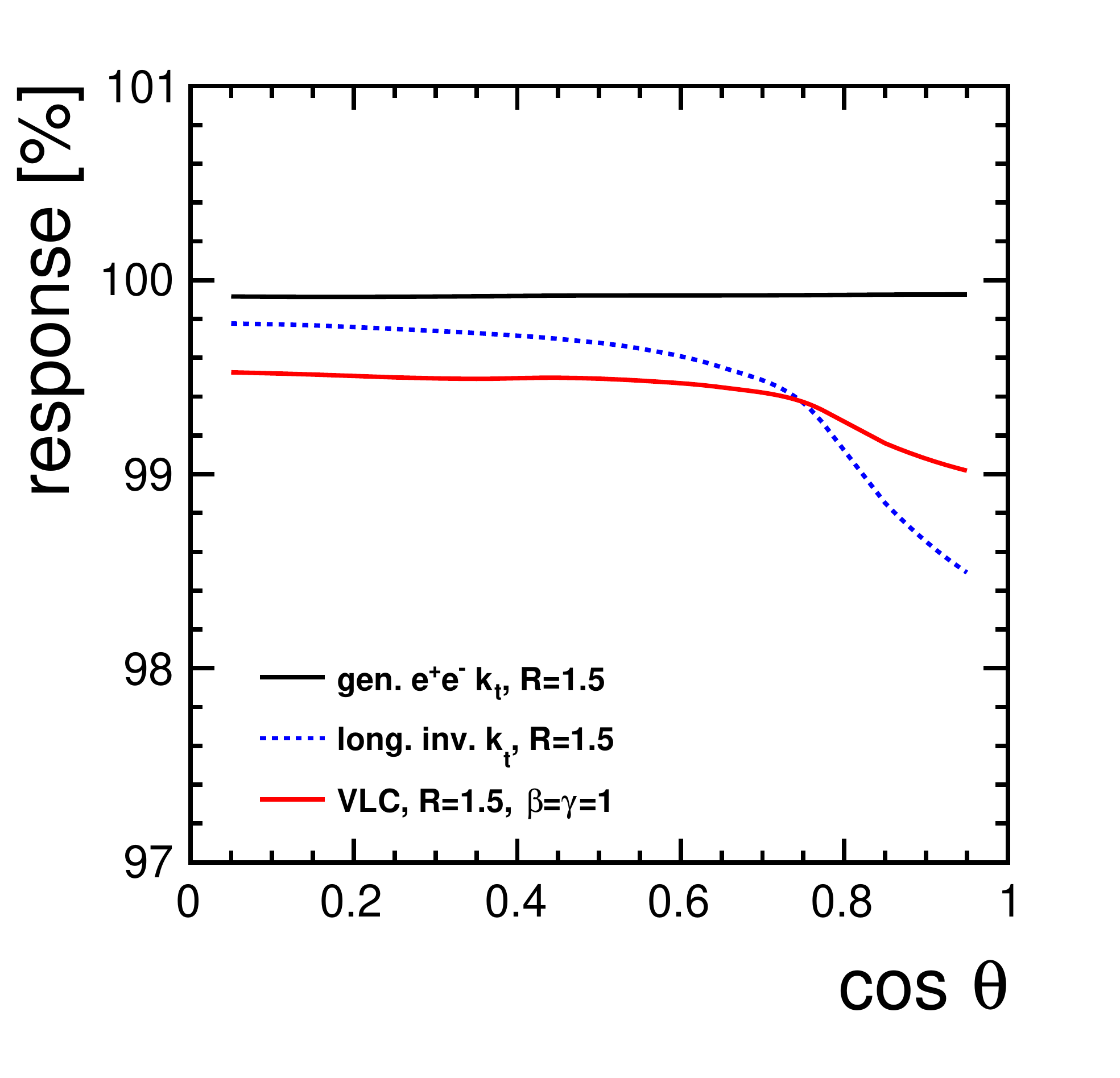}
\includegraphics[width=0.49\textwidth]{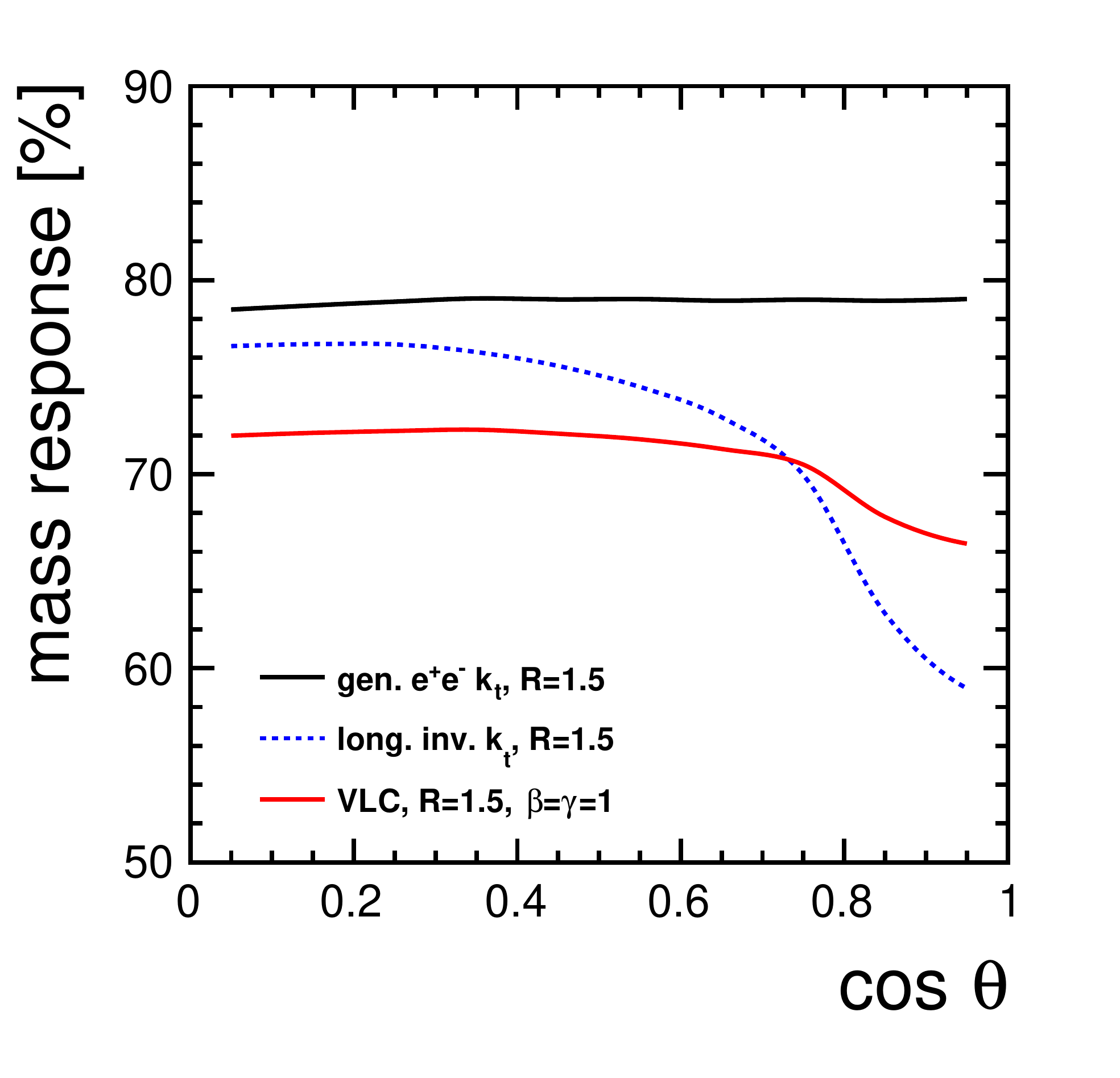}
\caption{The response to 1.5~\tev{} top jets as a function of polar angle for three jet algorithms, all with radius parameter $R=$ 1.5. The left plot shows the median reconstructed jet energy, the right plot the mean jet mass. Both quantities are normalized to the response of the Durham algorithm: $E=$ 1.5~\tev, $m\sim$ 370~\gev. }
	\label{fig:response}
}
\end{figure}

In Figure~\ref{fig:response} the energy and mass response is shown 
as a function of polar angle for three algorithms: 
the generalized $e^+e^-$ $k_t$ algorithm (black), the
longitudinally invariant $k_t$ algorithm (blue dashed) and VLC 
with $\beta=\gamma=$ 1 (red). The $R$-parameter is set to 1.5 for
all three algorithms. The generalized $e^+e^-$ $k_t$ algorithm
recovers over 99.9\% of the top quark energy for $R=$1.5,
independent of the jet polar angle. The shrinking
jet areas in the forward region of the longitudinally invariant 
$k_t$ and VLC algorithms lead to a slightly smaller response for 
$|\cos{\theta}| > $ 0.6. The polar angle dependence of 
longitudinally invariant $k_t$ is more pronounced.

The mass response of all three algorithms is substantially lower than
for the Durham algorithm. 
The generalized $e^+e^-$ $k_t$ algorithm has a flat response at nearly
80\%. The VLC and longitudinally invariant $k_t$ algorithms display
the same pattern as for the energy response: VLC starts off with
a lower response in the central region, but the response is
much flatter versus polar angle.

\subsection{Jet energy response with background}

To gain insight in the performance
in a more realistic environment with background, we overlay two hundred
1~\gev{} particles on each signal event. The background distribution
is strongly peaked in the forward direction following an exponential 
distribution peaked at $ \theta = $ 0, an approximation to 
the $\gamma \gamma \rightarrow $ {\em hadrons} background in 
energy-frontier electron-positron colliders (a more realistic
simulation of this background follows in Section~\ref{subsec:mcsetup}).

\begin{figure}[htbp!]
{\centering 
\includegraphics[width=0.49\textwidth]{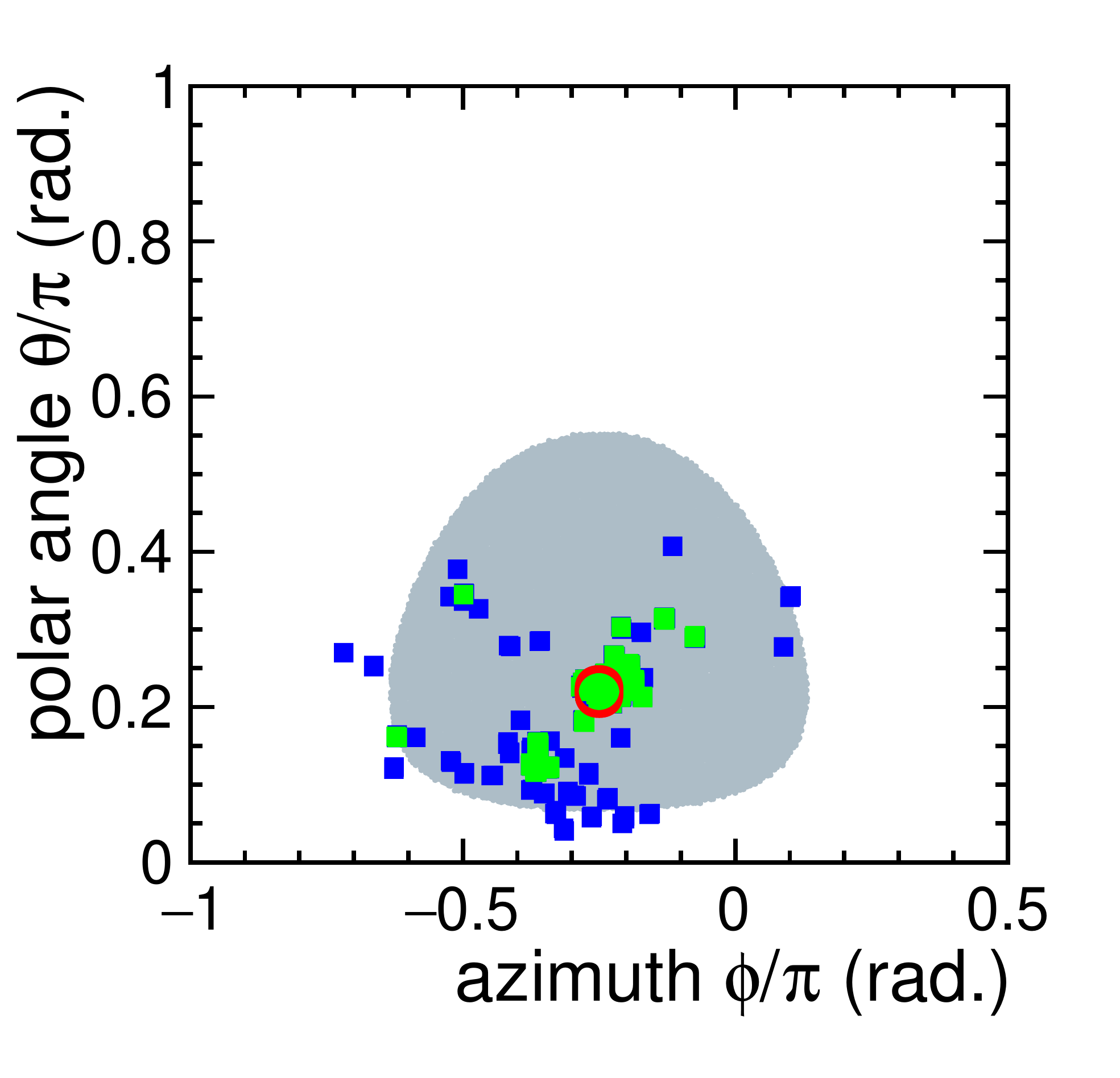}
\includegraphics[width=0.49\textwidth]{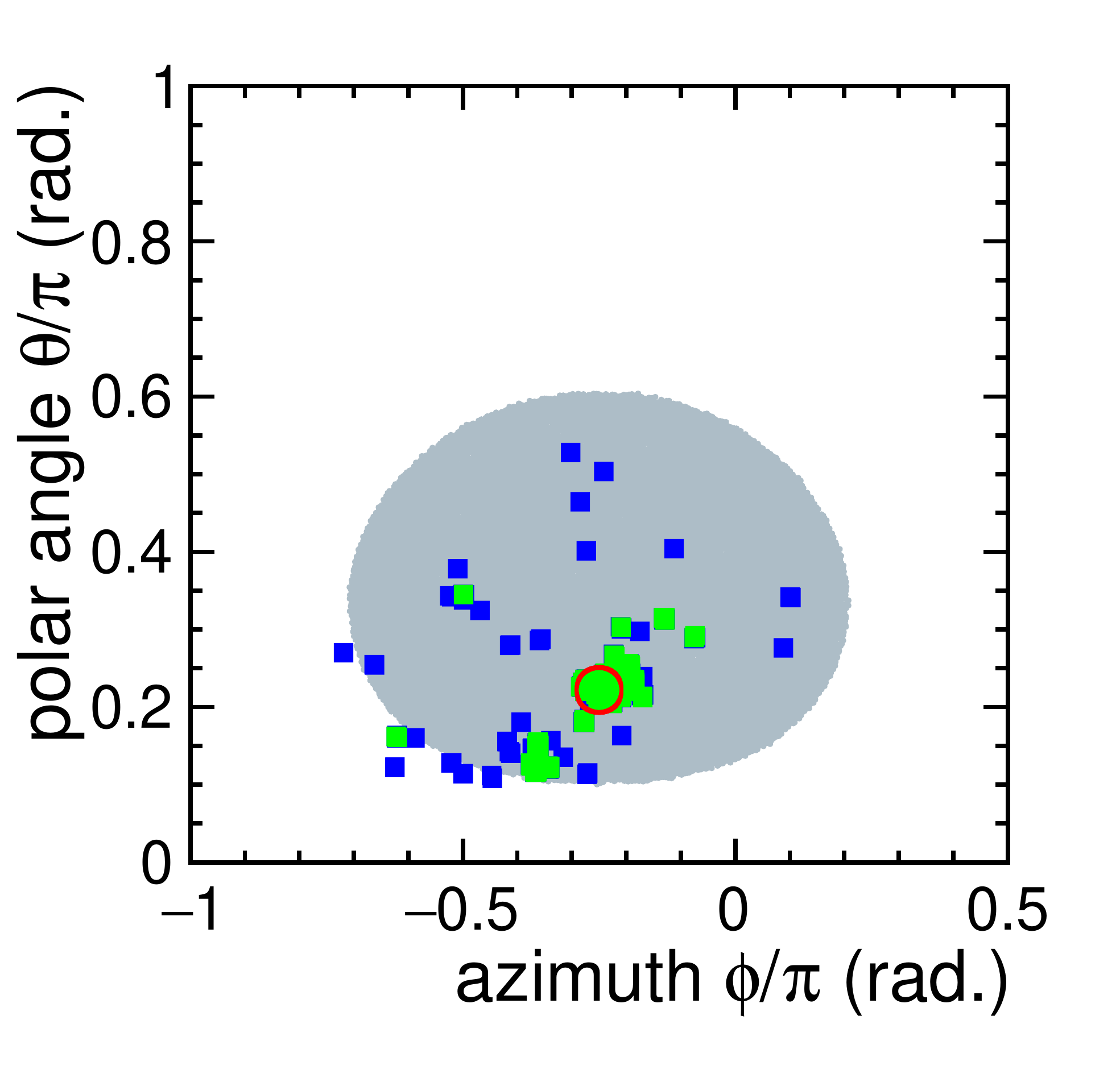}
\caption{Event display for a $e^+e^- \rightarrow t\bar{t}$ event at $\sqrt{s}=$ 3~\tev. The left panel shows the result of clustering with the longitudinally invariant $k_t$ algorithm with $R=$ 1.2, the right panel the corresponding VLC jet with the same radius parameter. The image zooms in on the $\theta - \phi$ area around one of the top jets. The location of the jet axis is indicated as a red circle. The area where the distance to the jet axis is smaller than the radius parameter ($\Delta R_{iC}$ for longitudinally invariant $k_t$, $ 1 - \cos \theta_{iC} $ for VLC) is indicated by the shaded region. The green squares represent particles from the top decay that are associated with each jet, the blue squares to
background particles clustered into the jet.  }
	\label{fig:display}
}
\end{figure}

The two event displays in Figure~\ref{fig:display} provide a zoom image
of the $\theta - \phi$ plane for a single event. The location of the 
jet axis is indicated as a red circle. The approximate catchment
area of both jets is shown in grey. The green squares represent particles 
from the top decay that are associated with each jet, the blue squares to
background particles clustered into the jet.  
Both algorithms find a very similar jet axis, centered on
the high-energy core of the jet. However, the algorithms have quite
distinctive footprints. The longitudinally invariant algorithms expose 
a larger area in the forward region, which renders it more vulnerable 
to background in this region.

\begin{figure}[htbp!]
{\centering 
\includegraphics[width=0.45\textwidth]{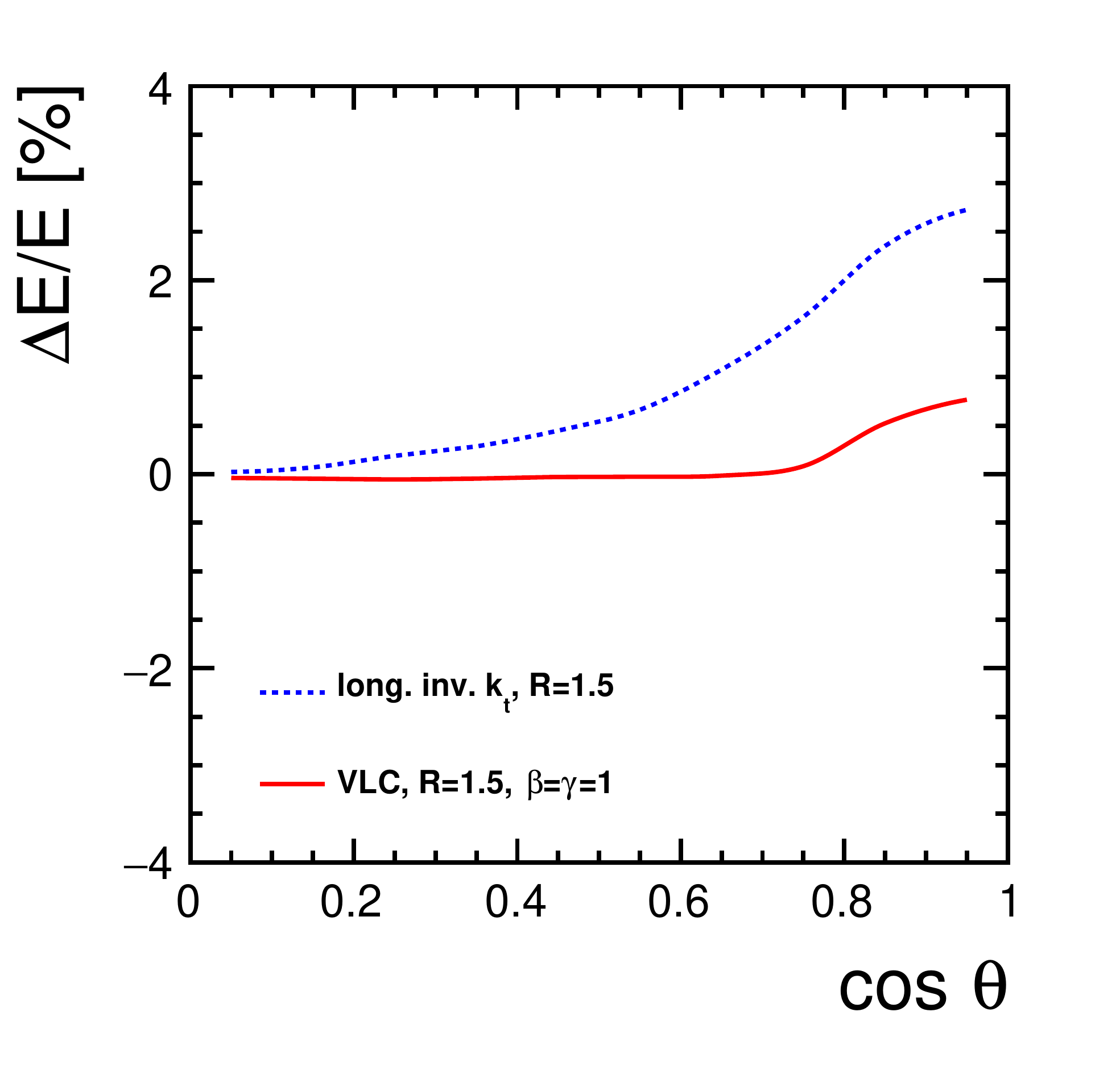}
\includegraphics[width=0.45\textwidth]{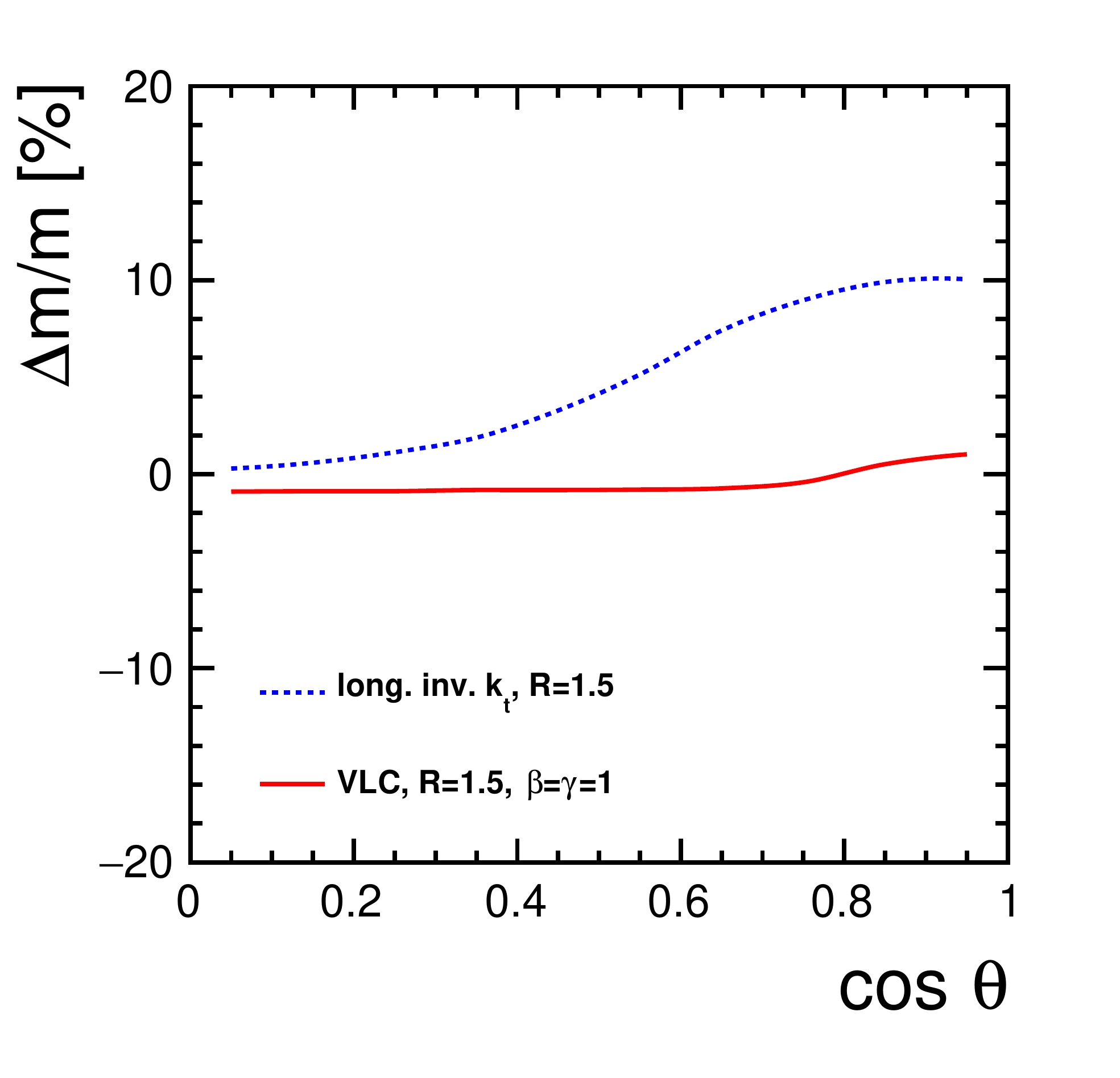}
\caption{The average contribution to the jet energy (left plot) and jet mass (right plot) of 200~\gev{} of forward-peaked background.}
	\label{fig:response_bkg}
}
\end{figure}

A quantitative view is obtained by comparing the energy and mass of
jets obtained when clustering the same events with and without background
particles. The bias (the average difference) in the jet energy and jet mass 
is shown in Figure~\ref{fig:response_bkg}. The background leads to 
a significant bias for forward jets reconstructed with the longitudinally
invariant $k_t$ algorithm. The VLC algorithm, on the other hand, is only
affected in the very forward region and the bias is much less pronounced.

The jet mass is known to be quite sensitive to soft and diffuse radiation, 
with the contribution scaling as the third power of the jet 
area~\cite{ATLAS:2012am}. We indeed find that the mass is strongly affected.
 A comparison of the jet reconstruction performance of the same process 
in a fully realistic environment is presented in 
Section~\ref{sec:resultstopclic3tev}.

\section{Results from full simulation}
\label{sec:simulation}

The performance of the different algorithms is compared in full-simulation
samples. We choose two benchmark scenarios
with fully hadronic final states that challenge jet reconstruction: 
di-Higgs boson production (with $h \rightarrow b\bar{b}$, i.e.\ 
a final state with four b-jets) and 
$ t \bar{t}$ production. Both analyses are performed 
at $\sqrt{s}= $3~\tev{}, with the CLIC\_ILD detector and
a realistic background of $\gamma \gamma \rightarrow ${\em hadrons}.

\subsection{Monte Carlo setup}
\label{subsec:mcsetup}

The studies in this Section are performed on CLIC 3~\tev{} Monte Carlo samples.
Events are generated with WHIZARD~\cite{Kilian:2007gr} (version 1.95).
The response of the CLIC\_ILD detector~\cite{Behnke:2013lya}
is simulated with GEANT4~\cite{Agostinelli:2002hh}. 
Multi-peripheral $\gamma \gamma \rightarrow ${\it hadrons} events are
generated with Pythia and superposed as {\em pile-up} on the signal events. 

At CLIC bunches are spaced by 0.5 ns and detector systems are 
expected to integrate the background of a number of 
subsequent bunch crossings. In this study, the background corresponding to 
60 bunch crossings is overlaid. In the event reconstruction, the information 
of the tracking system 
and the calorimeters is combined to form particle-flow objects (PFO) with the
Pandora~\cite{Marshall:2012hh} algorithm. 
Timing cuts on PFOs reduce the background level, with a very small impact on
the signal energy flow. The {\em nominal} (or {\em default}) selection of 
Ref.~\cite{Marshall:2012ry,Linssen:2012hp} reduces the 19~\tev{} 
of energy deposited in the calorimeters by the entire bunch train
to approximately 200~\gev{} superposed on a reconstructed events.
A more stringent set of cuts, referred to as the {\em tight} selection 
in Ref.~\cite{Linssen:2012hp}, reduces the background energy by
another factor of two. Both scenarios are studied in the following. 

The event simulation and reconstruction of the large data
samples used in this study was performed using the 
{\sc ILCDIRAC}~\cite{Grefe:2014sca,Tsaregorodtsev:2008zz} grid production tools.

\subsection{Higgs pair production}
\label{sec:resultsdihiggsclic3tev}

The study of Higgs boson pair production is crucial to assess the strength 
of the Higgs self-coupling. The analysis is very challenging at both
hadron and lepton colliders due to the very small cross section.
At an $e^+ e^-$ collider, the significance of this signal
is enhanced at large centre-of-mass energy, as the production rate 
in the vector-boson-fusion channel $e^+ e^- \rightarrow \nu \bar{\nu} hh$
grows strongly with centre-of-mass energy. In this section we focus on
events where both Higgs bosons decay to hadrons, through the dominant 
$h \rightarrow b \bar{b}$ decay of the SM Higgs boson. This final state
can be isolated~\cite{clichiggspaper} provided the four jets are reconstructed
with excellent energy resolution. 

The challenge of this measurement lies in the fact that both Higgs bosons 
are typically emitted at small polar angle~\cite{Fuster:2009em}. The most 
frequently observed topology has both Higgs bosons emitted in opposite 
directions: one in the 
forward direction and the other in the backward direction. At 3~\tev{} 
(at least) one of the Higgs bosons is emitted with $|\cos{\theta}| >$ 0.9 in 
approximately 85\% of events. In this area of the detector the background
level due to $\gamma \gamma \rightarrow$ {\em hadrons} production is
most prominent.
 
Despite the large centre-of-mass energy, the Higgs bosons are produced
with rather moderate energy: in 3~\tev{} collisions the most probable energy
of the Higgs bosons is approximately 200~\gev{}, with a long, scarcely
populated tail extending to 1.5~\tev. The modest Higgs boost is
sufficient for the b-quarks to continue in the same hemisphere 
as their parent Higgs boson, but it is rarely large enough 
for the Higgs boson to form a single jet.

\begin{figure}[t!] 
{\centering 
 \includegraphics[width=0.49\textwidth]{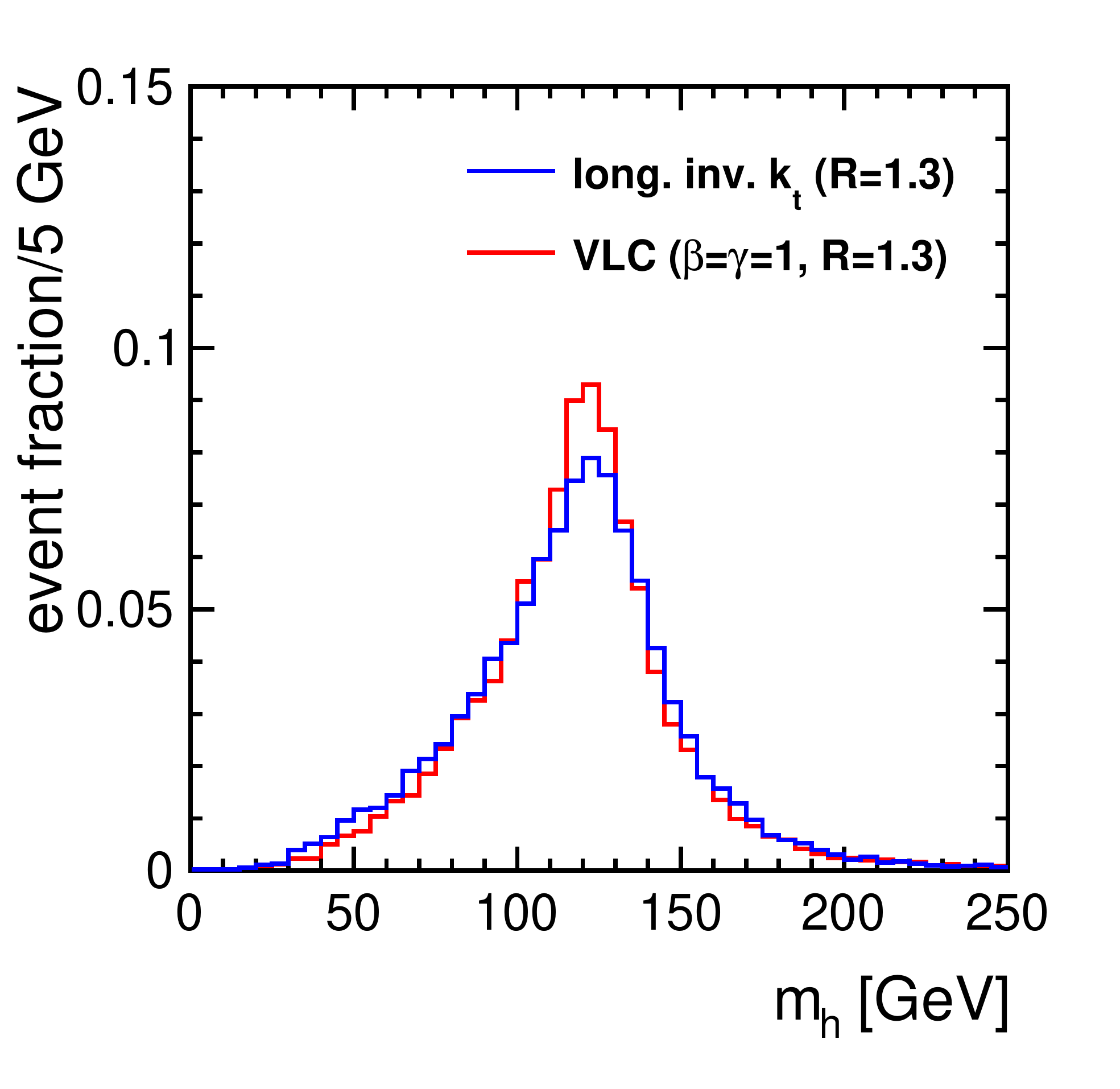} 
 \includegraphics[width=0.49\textwidth]{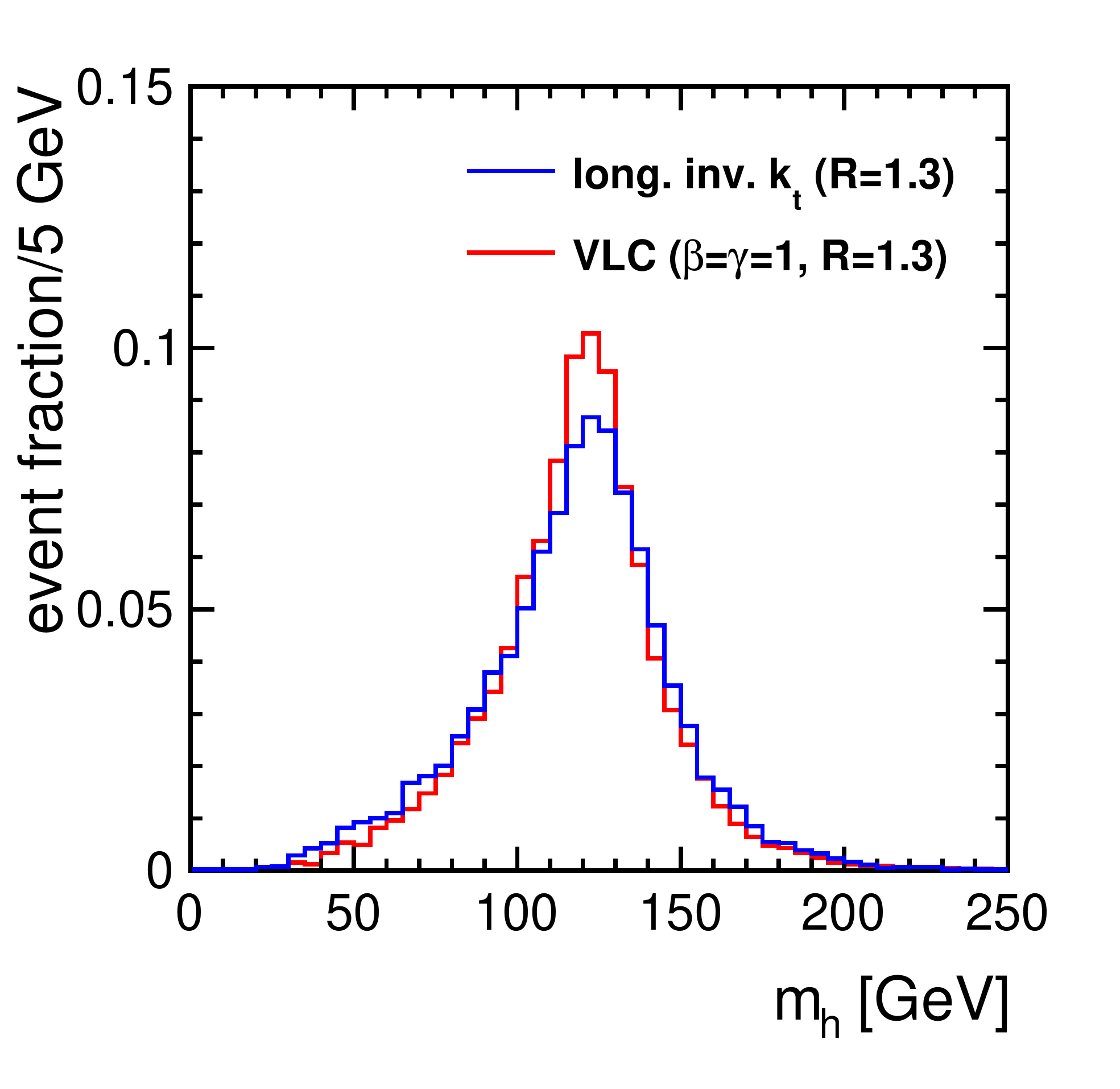}
\caption{The reconstructed di-jet mass distribution for fully hadronic decays of $e^+ e^- \rightarrow \nu \bar{\nu} hh$, $h\rightarrow b\bar{b}$ events at a 3~\tev{} CLIC. In the left panel all Higgs boson candidates are included, in the right panel only those that match onto exactly two b-quarks from Higgs boson decay. The nominal level of $\gamma \gamma \rightarrow$ {\em hadrons} background is overlaid on the signal. Particle flow objects are selected using the tight selection.}
  \label{fig:higgs_mass_clic}
}
\end{figure}

We perform exlusive jet reconstruction with $N_{jets}=$ 4. Higgs boson candidates
are reconstructed by pairing two out of the four jets. The combination
is retained that yields the best di-jet masses (i.e.\ that minimizes 
$\chi^2 = (m_{ij} - m_{h})^2 + (m_{kl} - m_{h})^2$, where $m_{ij}$ and $m_{kl}$
are the the masses of the two di-jet systems and $m_h = $ 126~\gev{} is 
the nominal Higgs boson mass used in the simulation). 

\begin{table}[b!]
        \begin{center}
	\caption{Response and resolution of the di-jet mass distributions obtained with the $k_t$ and VLC algorithms in the left panel of Figure~\ref{fig:higgs_mass_clic}. The columns list the median di-jet mass, the $\mathrm{IQR}_{34}$ and the $\mathrm{RMS}_{90}$. }
	\label{tab:results3tevhh}
	\begin{tabular}{l c c c}
		\hline
algorithm                         &   median [\gev] & $\mathrm{IQR}_{34}$ [\gev] &  RMS$_{90}$ [\gev] \\ \hline
$k_t$ ($R= $ 1.3)                 & 118.5 &  30.1 & 24.3 \\  
VLC ($R= $ 1.3, $\beta=\gamma=$ 1) & 118.9 & 27.1 & 22.0 \\ \hline
	\end{tabular}
	\end{center}
\end{table}

The distribution of the reconstructed mass of both di-jet systems 
forming the Higgs boson 
candidates is shown in Figure~\ref{fig:higgs_mass_clic}. The results
of two algorithms are shown, both with the radius parameter $R$ set to 1.3. 
The red line denotes the result of the VLC algorithm with 
$\beta = \gamma =$ 1, the black line that of the longitudinally 
invariant $k_t$ algorithm. Numerical results of the centre and width
of the reconstructed di-jet mass distribution are presented in 
Table~\ref{tab:results3tevhh}. 
The response of both algorithms is found to agree to within 0.5\%,
for all methods to estimate the central value of the distribution.
The Higgs mass resolution obtained with the VLC algorithm is  
better for both figures of merit. The $\mathrm{IQR}_{34}$ divided by the median
yields 22.6\% for the VLC algorithm versus 25.4\% for $k_t$.

%

\begin{figure}[t!] 
{\centering 
 \includegraphics[width=0.49\textwidth]{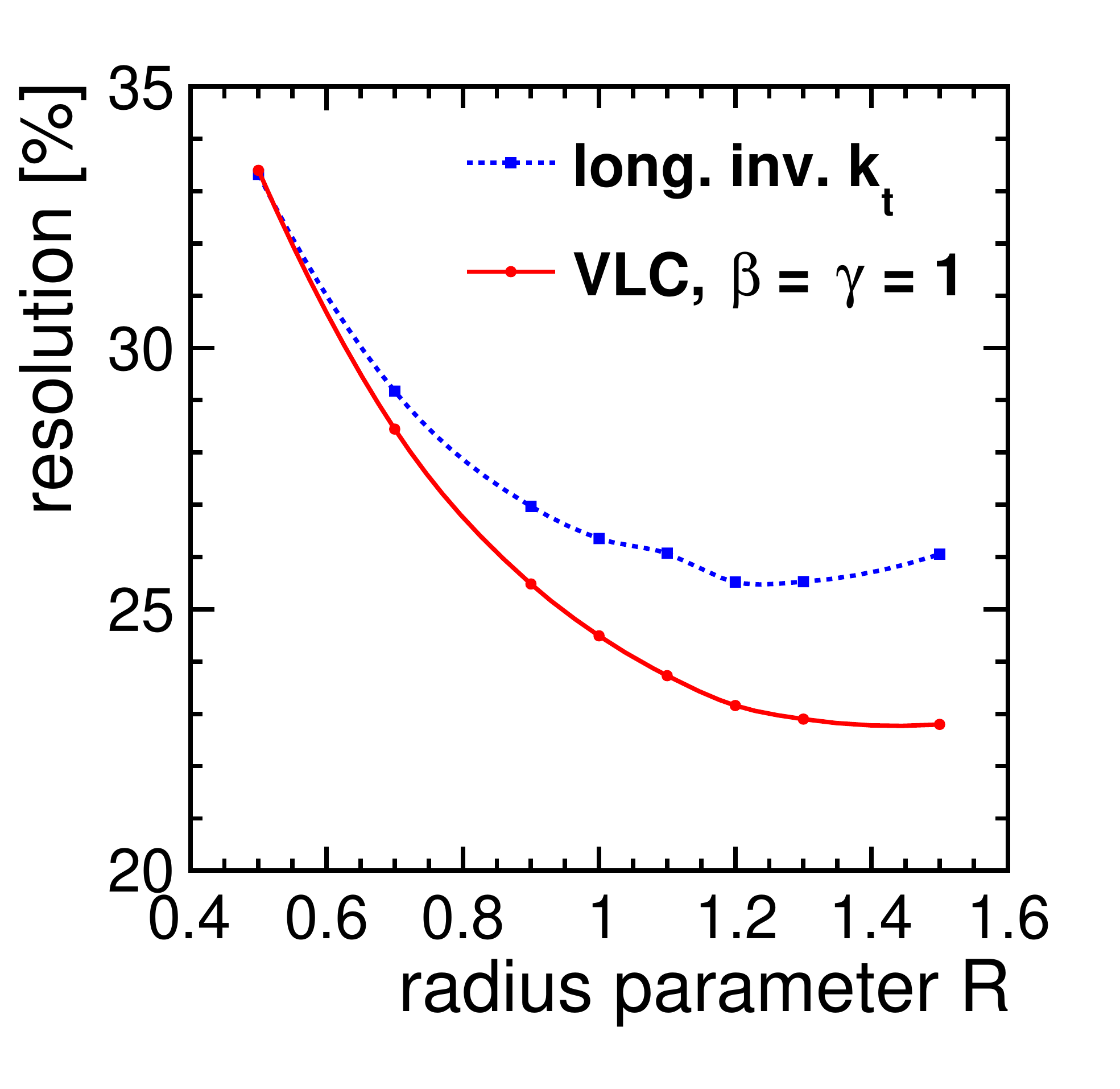} 
 \includegraphics[width=0.49\textwidth]{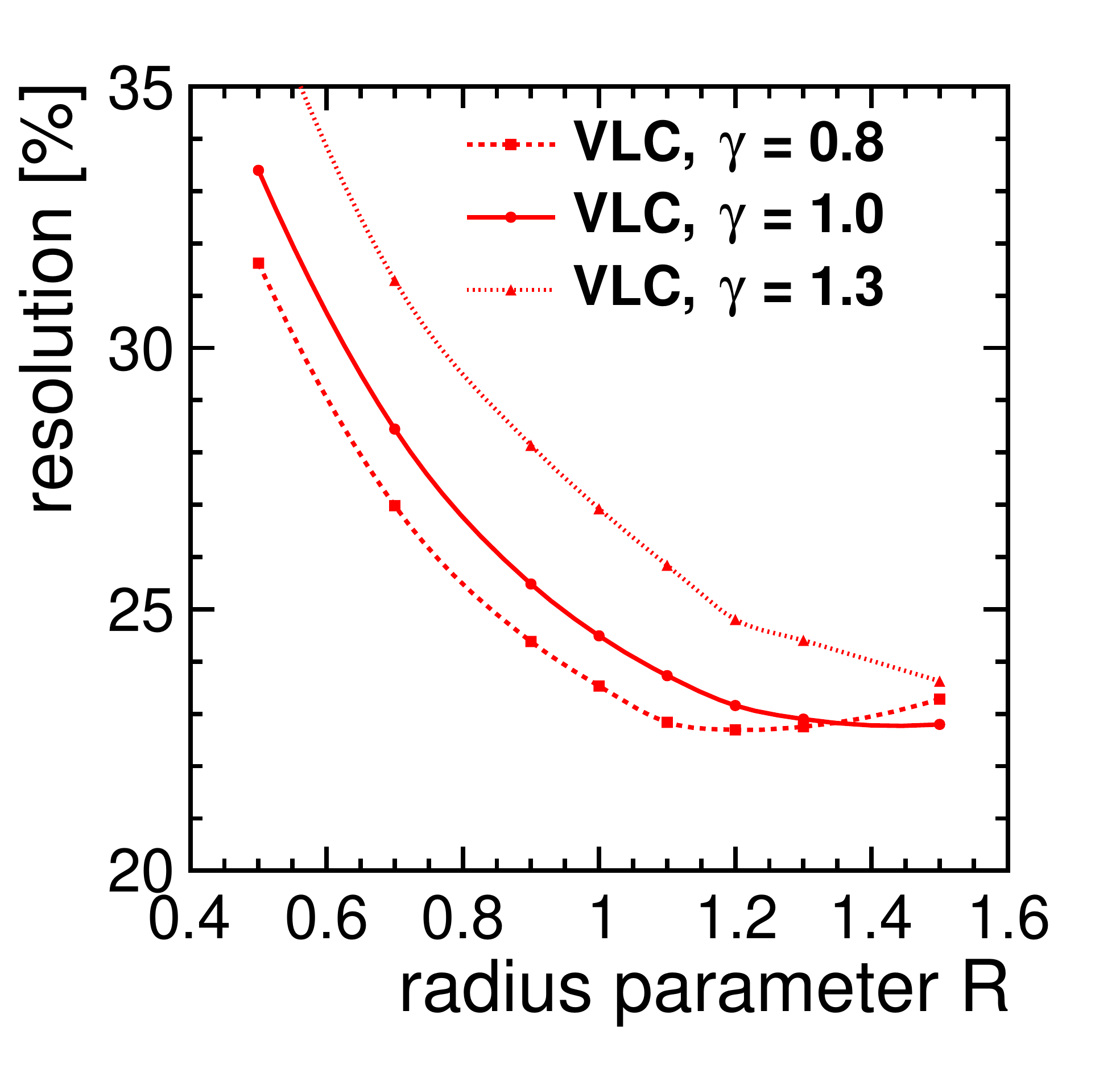}
\caption{The reconstructed di-jet mass resolution (determined as the 34\% inter-quantile range $\mathrm{IQR}_{34}$) for simulated, fully hadronic decays of $e^+ e^- \rightarrow \nu \bar{\nu} hh$, $h\rightarrow b\bar{b}$ events produced in 3~\tev{} $e^+ e^-$ collisions at CLIC. The nominal $\gamma \gamma \rightarrow$ {\em hadrons} background is overlaid on the signal event. Particle flow objects are selected using the tight selection.}
  \label{fig:higgs_clic_resolution_vs_r}
}
\end{figure}

The dependence of the $\mathrm{IQR}_{34}$ resolution on the radius parameter and the parameter $\gamma$ of the VLC algorithm is shown in Fig.~\ref{fig:higgs_clic_resolution_vs_r}. The best mass resolution is obtained for large values of
$R$ in both algorithms. The choice of $R \sim $ 1.3 is close to optimal
for both algorithms. Variation of the $ \gamma$ parameter,
that controls the evolution of the VLC jet area in the forward region, 
leads to a shift of the optimal value of $R$. With $\gamma<$ 1 the jet area 
is reduced at a slightly slower rate and the best resolution is obtained 
for smaller $R$. Choosing $\gamma>$ 1 the jet area shrinks more rapidly
and a larger $R$ is required to capture the complete energy flow.

\subsection{Top quark pair production}
\label{sec:resultstopclic3tev}
\begin{figure}[t!] 
{\centering 
 \includegraphics[width=0.49\textwidth]{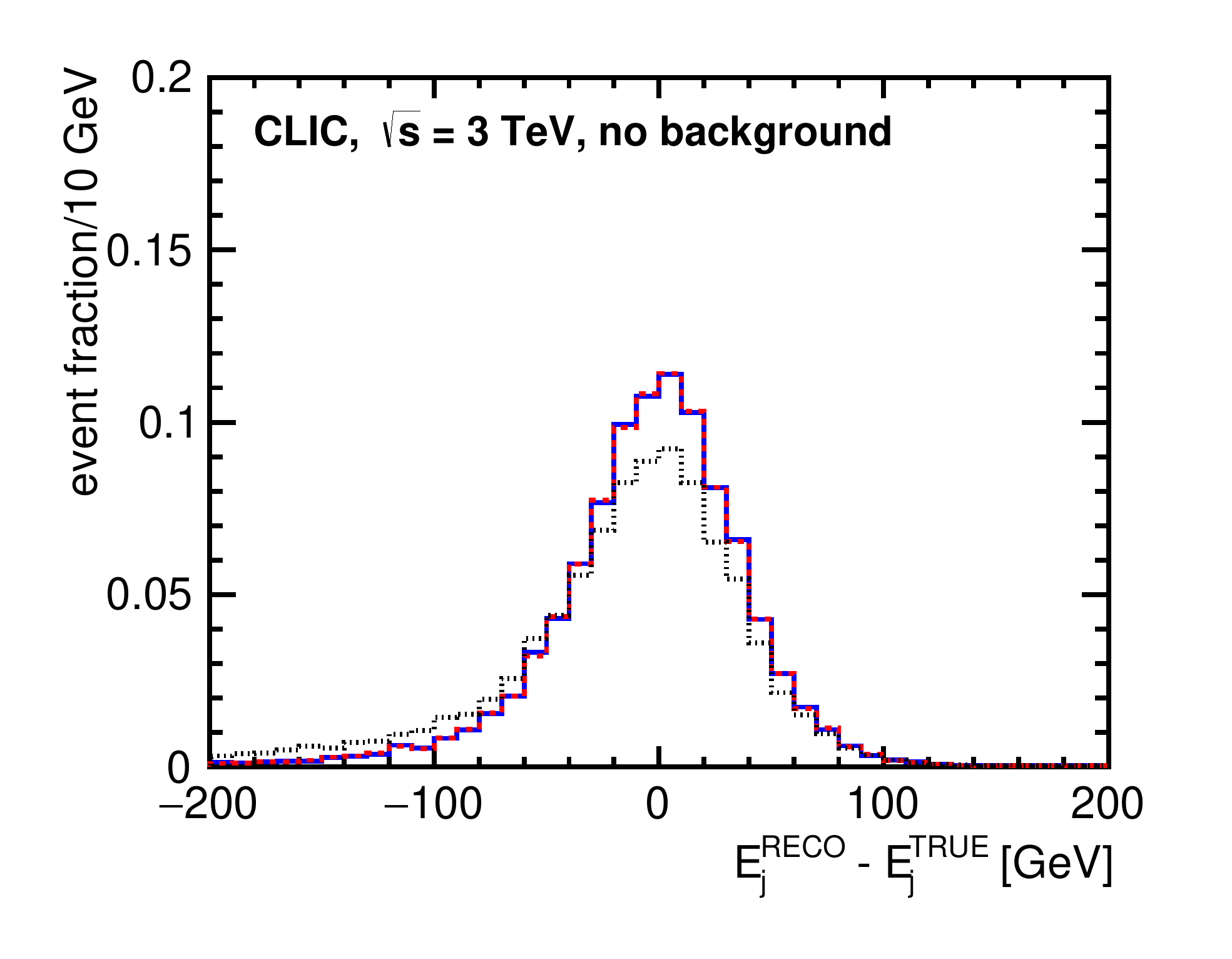} 
  \includegraphics[width=0.49\textwidth]{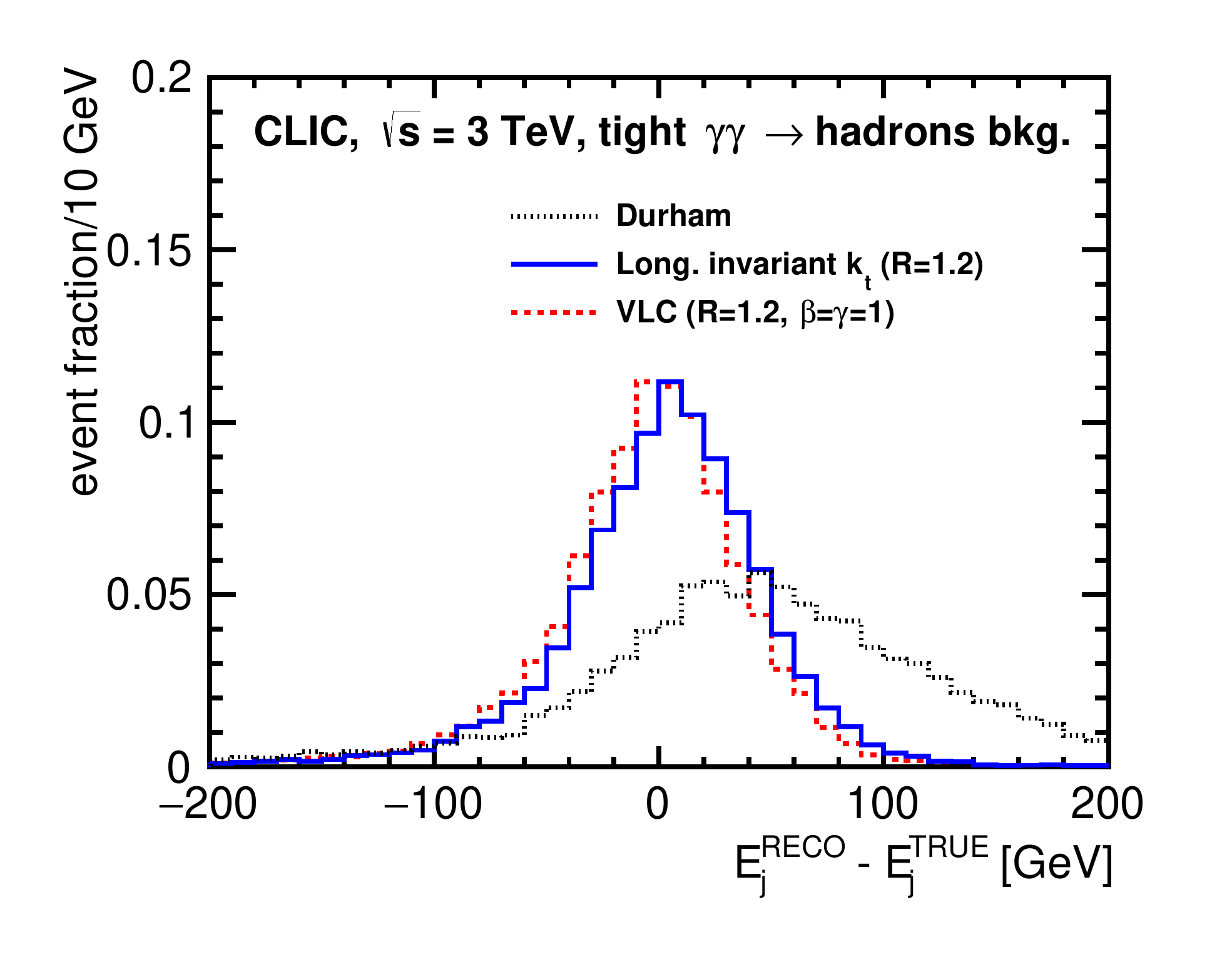} 

  \caption{The jet energy residuals (reconstructed minus true energy) for fully hadronic decays of $t\bar{t} $ events at a 3~\tev{} CLIC. No backgrounds are added in the left plot. In the right plot 60 bunch crossings of $\gamma \gamma \rightarrow$ {\em hadrons} background are overlaid on the signal and particle flow objects are selected using the tight selection.}
  \label{fig:clic_top_energy}
}
\end{figure}

\begin{figure}[t!] 
{\centering 
 \includegraphics[width=0.49\textwidth]{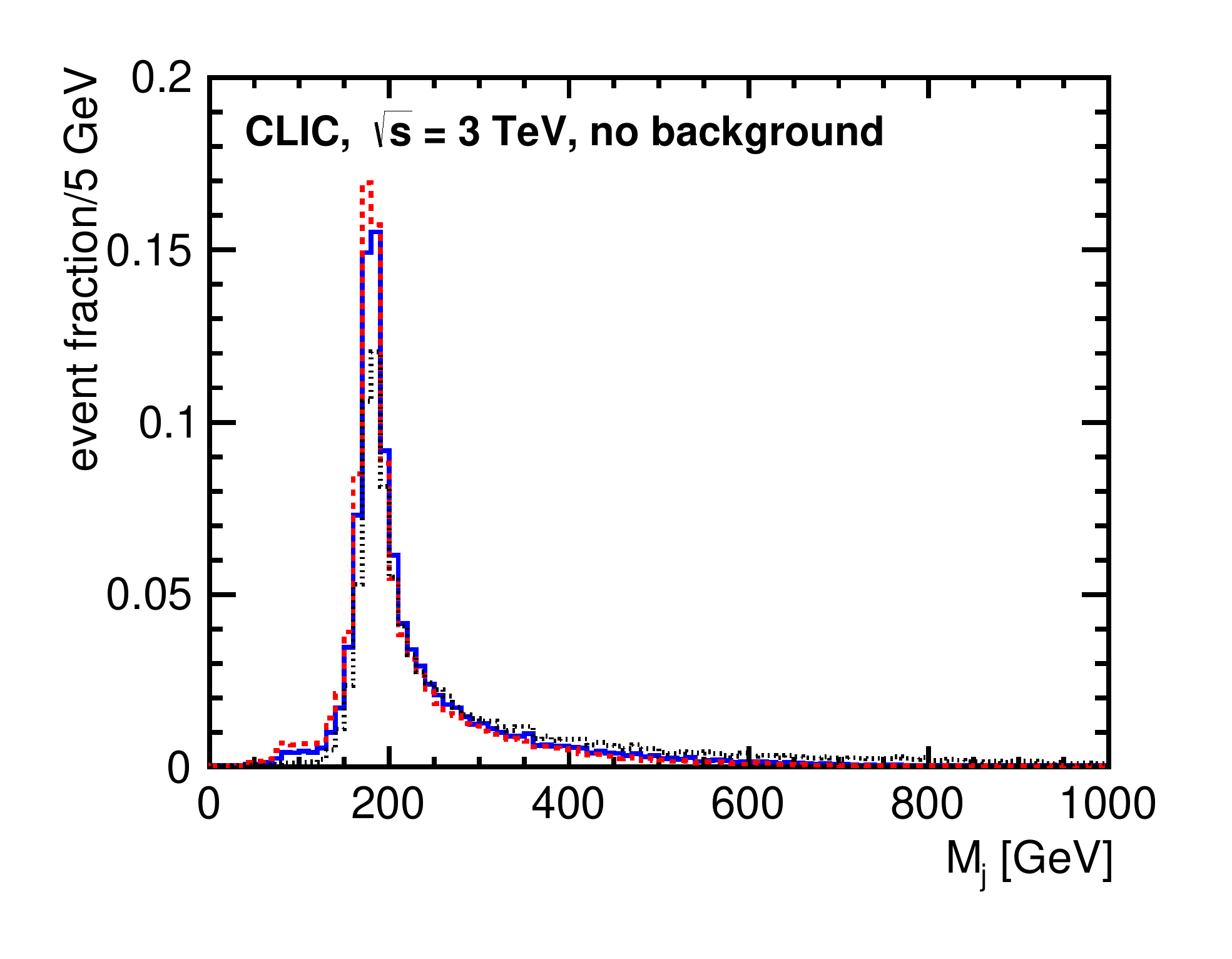} 
  \includegraphics[width=0.49\textwidth]{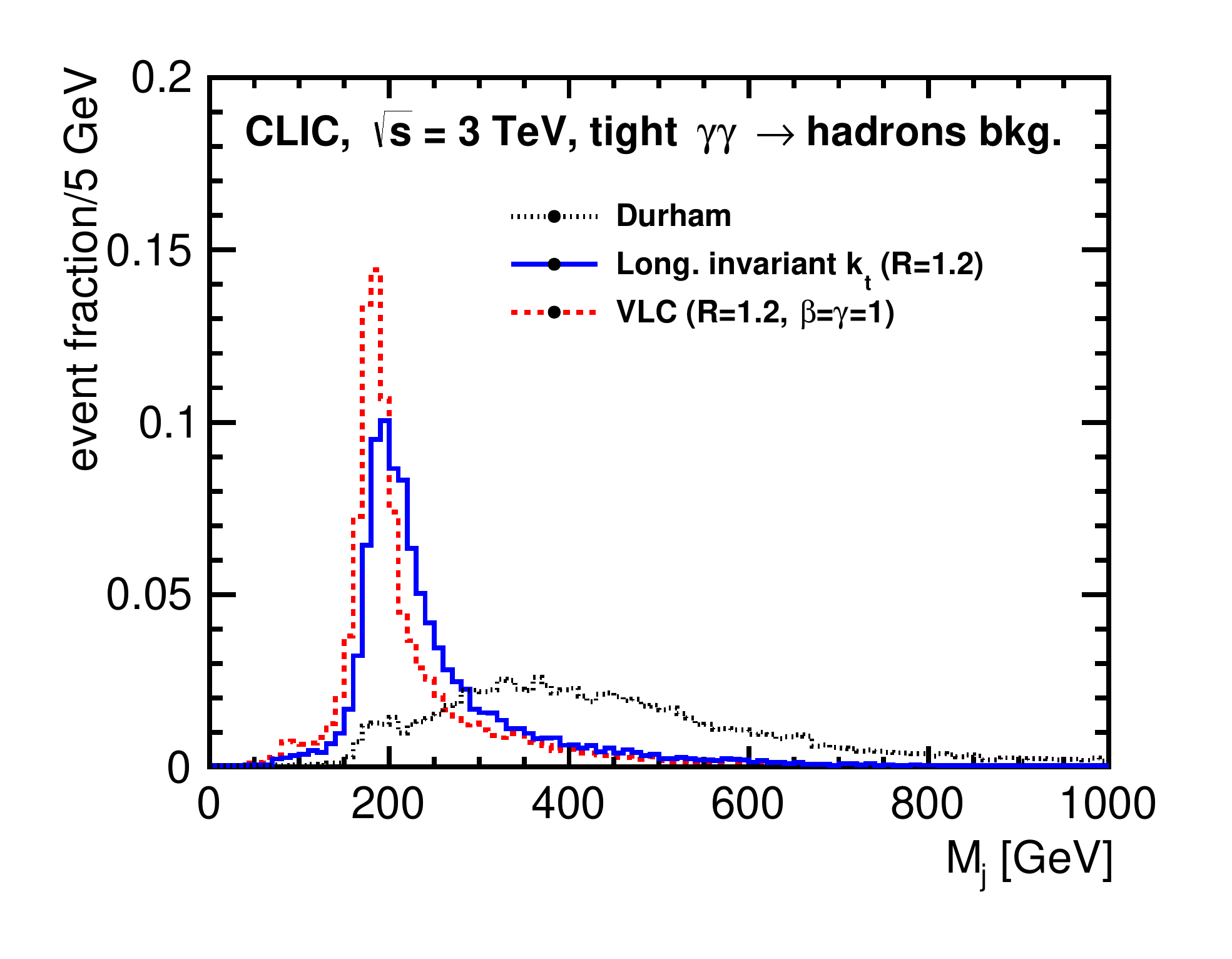} \\
  \caption{The reconstructed jet mass distribution for fully hadronic decays of $t\bar{t} $ events at a 3~\tev{} CLIC. No backgrounds are added in the left plot. In the right plot 60 bunch crossings of $\gamma \gamma \rightarrow$ {\em hadrons} background are overlaid on the signal and particle flow objects are selected using the tight selection.}
  \label{fig:clic_top_mass}
}
\end{figure}










The second benchmark we analyze is pair production of boosted top
quarks in multi-\tev{} operation of the CLIC $e^+e^-$ collider. At these energies the top quark 
decay products are so collimated that hadronic top quarks can be reconstructed 
as a single large-$R$ top-jet ($R\sim$1). Only the fully hadronic final state 
$e^+ e^- \rightarrow t \bar{t} \rightarrow b\bar{b}q\bar{q}'q''\bar{q}'''$ is 
considered. Events where either the top or anti-top quark is emitted in the 
forward or backward direction($|\cos{\theta}| >$ 0.7) are discarded 
to avoid the incomplete acceptance in that region.
To cope with the increased background at 3~\tev{} the {\em tight} PFO 
selection is applied. Jets are reconstructed with exclusive ($N=$2)
clustering with $R=$1.2. For comparison the same algorithm is also run on 
all stable Monte Carlo particles. These include neutrinos, but 
not the particles from $\gamma \gamma \rightarrow $ {\em hadrons} background.

The jet energy is a fairly good measure of the top quark energy.
The correction to the top quark energy is typically 
3.5\% and the energy resolution is typically 8\%. 
To measure the performance we compare the
jet reconstructed from particle flow objects with the jet found
by the same algorithm on the stable particles from the signal event
(i.e.\ excluding the $\gamma \gamma \rightarrow $ {\em hadrons}).
The jet energy residual is defined as the difference of the energy of 
detector-level and particle-level jets. The distribution is shown in 
Fig.~\ref{fig:clic_top_energy}.
The response is measured as the median of the residual distribution. 
The resolution is measured as the $\mathrm{IQR}_{34}$. 

\begin{table}[h!]
        \begin{center}
	\caption{The bias and resolution of the energy and mass measurements of reconstructed top jets in top quark pair production with fully hadronic top quark decay at a centre-of-mass energy of 3~\tev. Results are presented for the median response and two estimates of the resolution: the 34\% inter-quantile range ($\mathrm{IQR}_{34}$) and the RMS of central 90\% of jets ($\mathrm{RMS}_{90}$). All results are obtained by comparing the jet energy reconstructed from particle flow objects to the jet of stable MC particles from the signal event. The performance of the classical $e^+ e^-$ algorithm is such that the figures-of-merit cannot be estimated reliably under {\em nominal} background conditions (indicated by ``-'' entries in the Table).}
	\label{tab:results3tevtop}
	\begin{tabular}{l | ccc | ccc | ccc}
		\hline
\multicolumn{10}{c}{CLIC, $\sqrt{s} = $ 3~\tev{}, energy resolution  (no bkg./tight/nominal) $[\%]$ } \\ \hline
                    & \multicolumn{3}{|c}{median}   & \multicolumn{3}{|c}{$\mathrm{IQR}_{34}$} & \multicolumn{3}{|c}{$\mathrm{RMS}_{90}$} \\ \hline
Durham                 & -0.9 & 3.1 & -     & 4.6 & 6.6 & - & 3.7 & 5.7 & -   \\
generic $e^+ e^- k_t$ ($R=$1)   & -0.3 & 0.5 & -      & 3.4 & 4.0 & - & 2.7 & 3.4 & - \\
long. inv. $k_t$  ($R=$1.2)     & -0.2 & 0.4 & 1.8  & 3.1 &  3.2 & 3.4  & 2.5 & 2.7 & 2.8 \\
VLC              ($R=$1.2)      & -0.2 & -0.2 & 0.5 &  3.1 & 3.2 & 3.2 & 2.5 & 2.6 & 2.6 \\ \hline   
\multicolumn{10}{c}{CLIC, $\sqrt{s} = $ 3~\tev{}, mass resolution  (no bkg./tight/nominal) $[\%]$ } \\ \hline
                    & \multicolumn{3}{|c}{median}   & \multicolumn{3}{|c}{$\mathrm{IQR}_{34}$} & \multicolumn{3}{|c}{$\mathrm{RMS}_{90}$} \\ \hline
Durham                 & -1.0 & 37.7 & -     & 14.3 & - & - & 11.7 & 33.8 & -   \\
generic $e^+ e^- k_t$ ($R=$1)   & 0.5 & 4.7 & -      &  5.1 & 23.2 & - & 4.6 & 17.0 & - \\
long. inv. $k_t$    ($R=$1.2)   & 1.1 & 8.0 & 21.2  &  4.1 & 12.0  & 20.6 &  3.5 & 9.9 & 16.3 \\
VLC              ($R=$1.2)       & 0.8 & 1.7  & 5.6  &   4.1 & 7.1 & 9.4  & 3.5& 6.0 & 8.0 \\ \hline 
	\end{tabular}
	\end{center}
\end{table}

Quantitative results are presented in Table~\ref{tab:results3tevtop}.
The $\mathrm{RMS}_{90}$ 
is also presented to facilitate comparison to other studies.
In the absence of background, all algorithms reconstruct the energy
of the jet quite precisely, with a bias of less than 1\% and a resolution
of 2-4\%. 
The performance of the classical $e^+ e^-$ algorithms is degraded 
as soon as the $\gamma \gamma \rightarrow $ {\em hadrons} background 
with tight PFO selection is added.
The VLC and longitudinally invariant $k_t$ algorithms show very little
performance degradation even with the nominal PFO selection.

\begin{figure}[t!] 
{\centering 
 \includegraphics[width=0.49\textwidth]{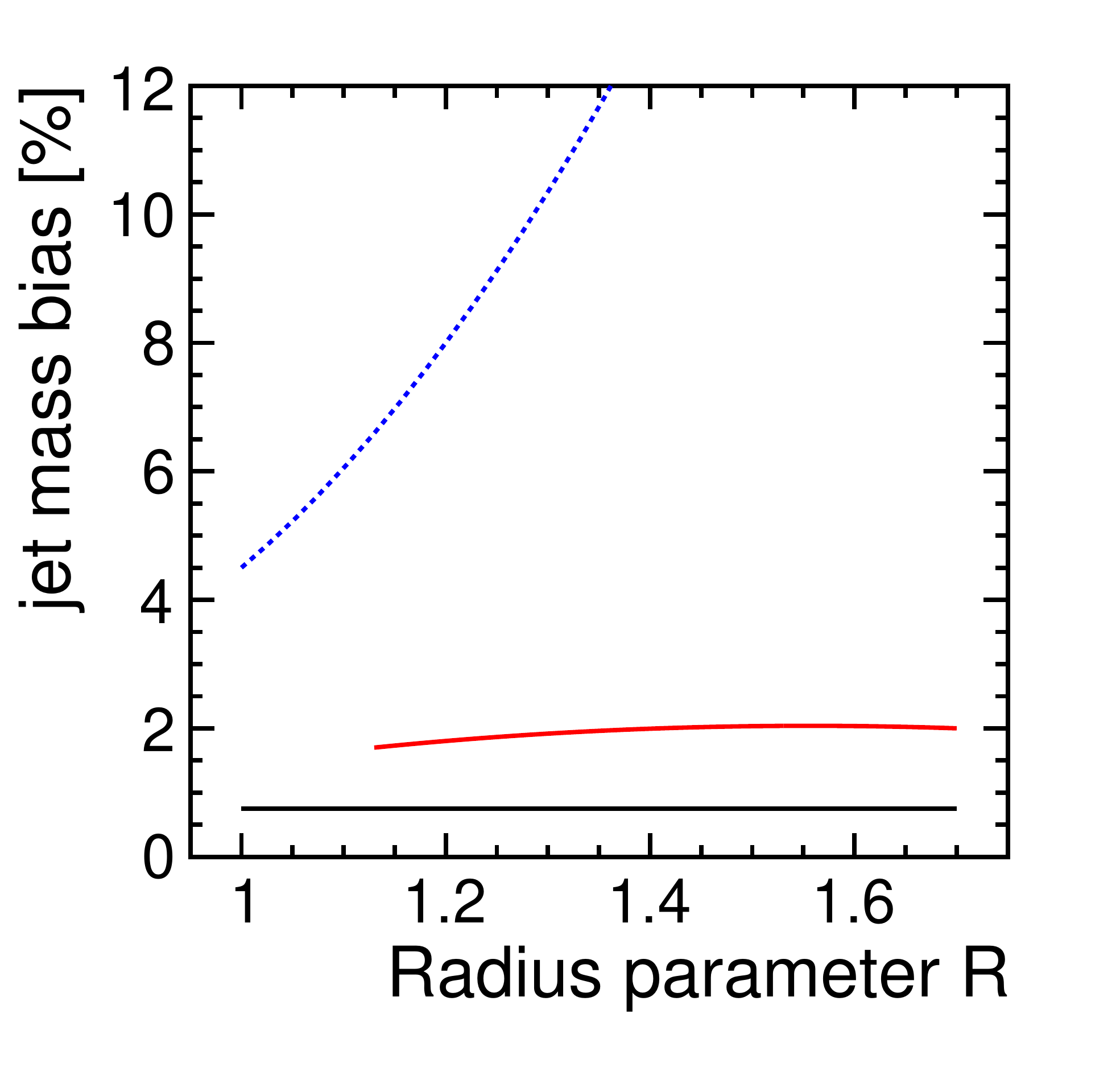} 
  \includegraphics[width=0.49\textwidth]{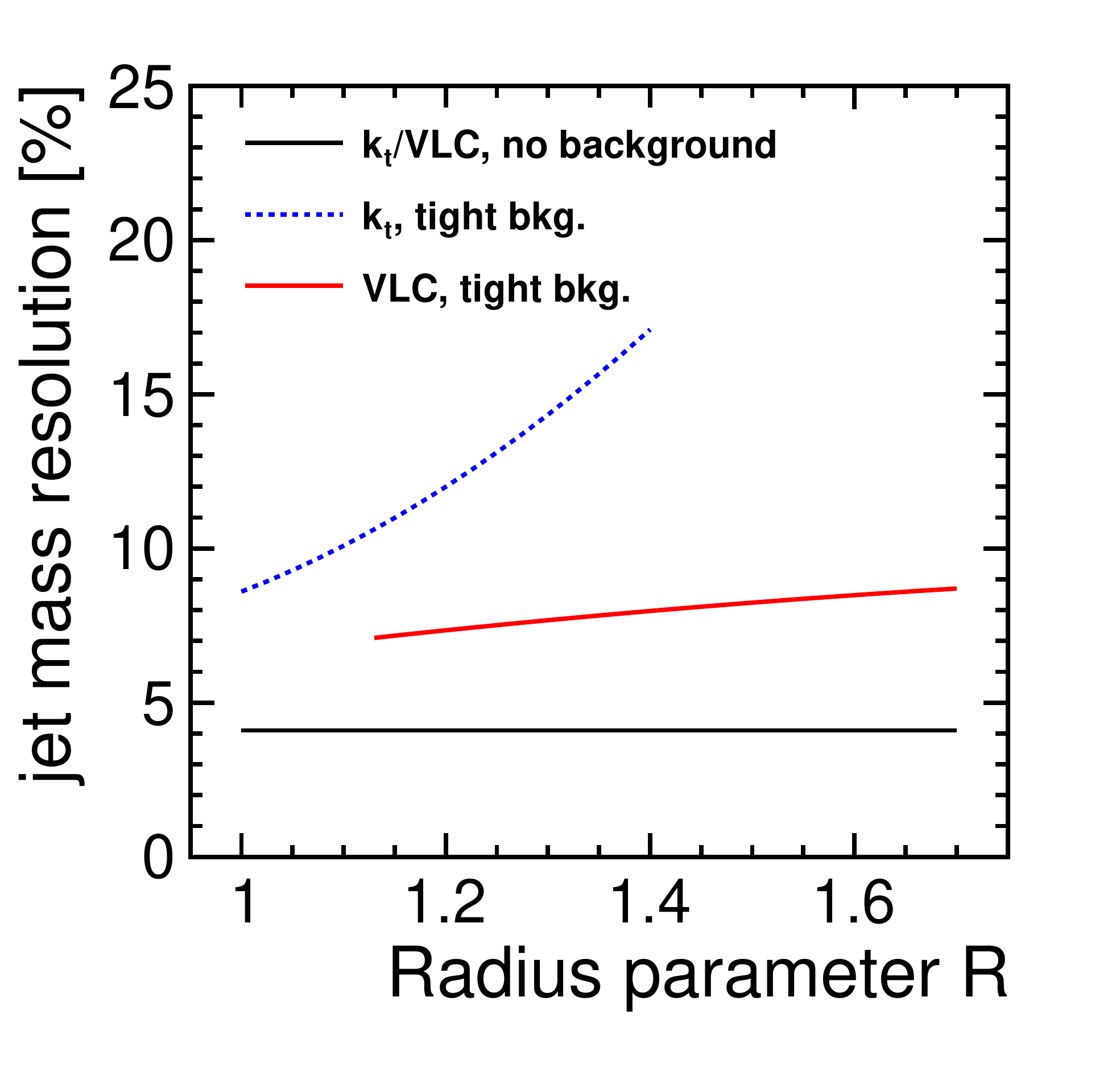} 
  \caption{The bias and resolution of the reconstructed jet mass versus radius parameter $R$ of two jet algorithms. The jets are reconstructed in fully hadronic top quark decays in $t\bar{t}$ events at a center-of-mass energy of 3~\tev{}. The jets reconstruced from particle-flow objects are compared to jets reconstructed from all stable particles from the signal event. The black curves correspond to the results obtained without background, the blue dashed and red curves to 60 bunch crossings of $\gamma \gamma \rightarrow ${\em hadrons} background overlaid on the signal, with the {\em tight} PFO selection.}
  \label{fig:toy_study}
}
\end{figure}

The jet invariant mass is much more sensitive to soft background
contamination~\cite{Dasgupta:2007wa,ATLAS:2012am}. 
The jet mass distributions are presented in 
Figure~\ref{fig:clic_top_mass}. In the left panel, which corresponds to
$t\bar{t}$ events without background overlay, all algorithms are seen
to reconstruct a narrow peak close to the top quark mass. The long tail 
toward large mass is due to radiation off the top quark and its decay 
products and is also present in the jets reconstructed from stable MC 
particles. The plots in the right panel show a severe degradation
when the  $\gamma \gamma \rightarrow $ {\em hadrons} background 
with tight PFO selection is added, most noticeably for the Durham algorihtm. 
The bias and resolution of
the jet mass is shown as a function of radius parameter in 
Fig.~\ref{fig:toy_study}.

A quantitative summary is presented 
in the second part of Table~\ref{tab:results3tevtop}.
The bias on the jet mass without background is sub-\% for most algorithms.
The resolution of the VLC and longitudinally invariant $k_t$ algorihms
is significantly better than that of the classical $e^+ e^-$ algorithms.
The 4.1\% resolution is a testimony to the potential of highly granular
calorimeters and particle flow reconstruction for jet substructure 
measurements.

The $\gamma \gamma \rightarrow $ {\em hadrons} background has a profound
effect on the performance. The performance of the
classical algorithms is clearly inadequate, with a strong bias and
a severe degradation, even with the tight PFO selection. 
The VLC and longitudinally invariant $k_t$ algorihms are much less
affected, as expected from the smaller exposed area.
The VLC algorithm is found to be more resilient than the longitudinally
invariant $k_t$, confirming the result anticipated at the particle level
in Section~\ref{sec:toyexamples}.

\section{Discussion}
\label{sec:discussion}

In this Section we discuss the implications for lower-energy colliders and
identify several topics that merit further study.

\subsection{Implications for lower-energy lepton colliders}

In this paper we have focused on CLIC operation at $\sqrt{s} = $ 3~\tev{},
arguably the most challenging environment that lepton colliders might
face in the next decades. We have chosen this environment 
because subtle differences in jet definitions 
lead to significant differences 
in performance. This has helped us to gain a deeper understanding of the 
intricacies of jet reconstruction at lepton colliders and to establish
solid conclusions about the resilience of the different algorithms.

These findings are by no means limited to the CLIC environment. Subtle but
significant differences in performance are expected also for the ILC
at 250~\gev{} or 500~\gev{} and at circular colliders.

\subsection{Further R\&D on jet algorithms}

The set of jet algorithms studied in this paper is by no means exhaustive.
This study does not address several recent proposals, such as the XCone 
algorithm~\cite{Stewart:2015waa} or the {\em global} jet clustering 
proposed by Georgi~\cite{Georgi:2014zwa}. 

A broad range of new techniques developed for the LHC have so far remained
unexplored. This is particularly true for a set of tools that has proven 
extremely powerful in pile-up mitigation and correction in ATLAS and CMS.

Jet grooming 
(the collective name for (mass-drop) filtering~\cite{Butterworth:2008iy}, 
pruning~\cite{Ellis:2009su} and trimming~\cite{Krohn:2009th}) effectively
reduces the exposed jet area to several small regions with large
energy flow. This provides an effective means of capturing a large fraction 
of the jet energy while reducing the impact of {\em soft} contamination. 
Tests of the trimming algorithm in the CLIC environment yield very good 
results, improving the jet mass resolution significantly. 

Techniques to correct for the effect of pile-up based on an event-by-event 
measurement of the pile-up activity~\cite{Cacciari:2007fd} are 
quite successful at the LHC. Subtraction at the constituent level 
with dynamical 
thresholds~\cite{Cacciari:2014gra,Bertolini:2014bba,Berta:2014eza} 
is under active development. An adaptation to the environment 
at lepton colliders, with a very sparse background energy flow,
may prove useful.

\section{Conclusions}
\label{sec:conclusions}

We have studied the jet reconstruction performance of several 
sequential reconstruction algorithms at high-energy lepton colliders.
In addition to the classical $e^+ e^-$ algorithms we include 
a version of the same algorithms with beam jets. We also study
the performance of the longitudinally invariant $k_t$ algorithm 
and a new $e^+ e^-$ algorithm, called VLC~\cite{Boronat:2014hva}, 
which are expected to be more resilient to the impact of backgrounds.

The study is based on detailed Monte Carlo simulation. For
two benchmark processes we use a full simulation of the linear
collider detector concepts, including the relevant background processes.

The perturbative energy corrections of all algorithms with finite size jets 
are sizeable for small values of the radius parameter (10-15\% for $R=$0.5)
and decrease to 1-5\% for $R=$ 1.5. This result is approximately independent
of the process and centre-of-mass energy. 
Convergence with $R$ is faster for the generalized $e^+ e^-$ algorithm
than for the longitudinally invariant $k_t$ algorithm and VLC, that 
expose a smaller area to forward jets.
  
The non-perturbative hadronization correction represents a very small part 
of the total correction. Its relevance decreases with increasing
centre-of-mass energy: it is less than 1\% at $\sqrt{s}=$ 250~\gev{}
for all algorithms and less than a per mille at $\sqrt{s}=$ 3~\tev{}.
The generalized $e^+ e^-$ algorithm again converges fastest.
We have estimated the non-perturbative correction also for the
jet invariant mass: these corrections are much larger than the 
non-perturbative energy corrections and remain of the order of a few \%
even for $R=$ 1.5 and $\sqrt{s}=$ 3~\tev{}.

The forward-peaked $\gamma \gamma \rightarrow ${\em hadrons} background 
at future high-energy linear lepton colilders is one of the most important
factors in the jet reconstruction performance. It has motivated
ILC and CLIC to abandon classical, inclusive algorithms in favour
of algorithms with a finite jet size. Algorithms that expose a reduced
solid angle in the forward region of the detector, such as 
longitudinally invariant $k_t$ or VLC are more robust. A particle-level
study shows that these two algorithms have a different response, with
VLC showing a lower response, but one that is more stable versus
polar angle. VLC is found to be less susceptible to background.

We present two studies in full simulation, namely
di-Higgs production and top quark pair production at $\sqrt{s} =$ 3~\tev,
which present a combination of a relatively harsh background level,
high jet multiplicity and forward jets. In both cases the classical
 $e^+ e^-$ algorithms offer an inadequate performance. The same is true
for the generalized version with beam jets. VLC provides significantly
better mass resolution for the Higgs study and considerably better
jet mass reconstruction than the longitudinally invariant 
$k_t$ algorithm.

Jet clustering is key technique for many analyses of multi-jet final states
at future high-energy electron-positron colliders. This study shows that 
a considerable increase in performance can be obtained by a careful choice
of the clustering algorithm. We recommend, therefore, that studies into the
physics potential of future $e^+ e^-$ colliders carefully optimize the choice 
of the jet reconstruction algorithm and its parameters.
We also encourage further work on robust algorithms for $e^+ e^-$ collisions.

\section*{Acknowledgement}
This work has been carried out in the framework of the CLICdp collaboration.
The authors acknowledge the effort of the ILC and CLIC detector \& physics 
groups in putting together the simulation infrastructure used to benchmark
the algorithms. This work benefited from services provided by the ILC Virtual 
Organisation, supported by the national resource providers of the EGI 
Federation. The authors of the VLC algorithm 
would like to thank Gavin Salam and Jesse Thaler for helpful 
suggestions and the FastJet team for guidance creating the plugin code.

\printbibliography[title=References]
	
\end{document}